\documentclass[5p,times]{elsarticle}

\usepackage{lipsum}
\usepackage{xcolor}
\usepackage{pdfpages}
\usepackage{amssymb}
\usepackage[utf8]{inputenc}
\usepackage{lineno,hyperref}

\usepackage{graphicx}
\graphicspath{ {./graphics/} }
\usepackage{color,soul}

\usepackage{pdflscape}
\usepackage{lscape}
\usepackage{afterpage}
\usepackage{caption}
\usepackage{longtable}
\usepackage[normalem]{ulem}
\useunder{\uline}{\ul}{}
\usepackage{lscape}
\usepackage{longtable}
\usepackage[flushleft]{threeparttable}
\usepackage{amsmath,graphicx,soul,subcaption,balance}

\usepackage{multirow}
\usepackage{multicol}
\usepackage{tabu}
\usepackage{tabularx}
\usepackage[export]{adjustbox}

\usepackage{booktabs}
\usepackage{array}
\usepackage{pdflscape}

\usepackage[superscript,biblabel]{cite}
\modulolinenumbers[5]

\journal{}

\begin{document}

\begin{frontmatter}

\title{Continental generalization of an {AI} system for clinical seizure recognition}

\author[affusyd1,affusyd2]{Yikai~Yang}
\ead{yikai.yang@sydney.edu.au}

\author[affusyd1,affusyd2]{Nhan~Duy~Truong}
\ead{duy.truong@sydney.edu.au}

\author[affusyd1,affusyd2]{Christina~Maher}
\ead{christina.maher@sydney.edu.au}

\author[affrpa1,affrpa2]{Armin~Nikpour}
\ead{armin@sydneyneurology.com.au}

\author[affusyd1,affusyd2]{Omid~Kavehei\corref{correspondingauthor}}
\cortext[correspondingauthor]{Corresponding author}
\ead{omid.kavehei@sydney.edu.au}

\address[affusyd1]{School of Biomedical Engineering, Faculty of Engineering, The University of Sydney, NSW 2006, Australia}
\address[affusyd2]{Australian Research Council Training Centre for Innovative BioEngineering, The University of Sydney, NSW 2006, Australia}
\address[affrpa1]{Comprehensive Epilepsy Service and Department of Neurology at the Royal Prince Alfred Hospital, NSW 2050}
\address[affrpa2]{Faculty of Medicine and Health, central clinical school, The University of Sydney, NSW 2006, Australia}

\begin{abstract}
{\bf Background}: Electroencephalogram (EEG) monitoring and objective seizure identification is an essential clinical investigation for some patients with epilepsy. Accurate annotation is done through a time-consuming process by EEG specialists. Computer-assisted systems for seizure detection currently lack extensive clinical utility due to retrospective, patient-specific, and/or irreproducible studies that result in low sensitivity or high false positives in clinical tests. We aim to significantly reduce the time and resources on data annotation by demonstrating a continental generalization of seizure detection that balances sensitivity and specificity.
\\ \\
{\bf Methods}: This is a prospective inference test of artificial intelligence on nearly 14,590 hours of adult EEG data from patients with epilepsy between 2011 and 2019 in a hospital in Sydney, Australia. The inference set includes patients with different types and frequencies of seizures across a wide range of ages and EEG recording hours. The artificial intelligence (AI) is a convolutional long short-term memory network that is trained on a USA-based dataset. The Australian set is about 16 times larger than the US training dataset with very long interictal periods (between seizures), which is way more realistic than the training set and makes our false positives highly reliable. We validated our inference model in an AI-assisted mode with a human expert arbiter and a result review panel of expert neurologists and EEG specialists on 66 sessions to demonstrate achievement of the same performance with over an order-of-magnitude reduction in time.
\\ \\
{\bf Findings}: Our inference on 1,006 EEG recording sessions on the Australian dataset achieved 76.68\% with nearly 56 [0, 115] false alarms per 24 hours on average, against legacy ground-truth annotations by human experts, conducted independently over nine years. Our pilot test of 66 sessions with a human arbiter, and reviewed ground truth by a panel of experts, confirmed an identical human performance of 92.19\% with an AI-assisted system, while the time requirements reduce significantly from 90 to 7.62 minutes on average.
\\ \\
{\bf Interpretation}: Accurate and objective seizure counting is an important factor in epilepsy. An AI-assisted system can help improve efficiency and accuracy alongside human experts, particularly in low and middle-income countries with limited expert human resources.
\\ \\
{\bf Fundings}: SOAR Fellowship from The University of Sydney, a Microsoft AI for Accessibility grant, and a Research Training Program (RTP) support provided by the Australian Government.
\end{abstract}
\end{frontmatter}

\noindent
\begin{figure*}[!ht]
\colorbox{purple!15}{
\begin{minipage}[t]{0.97\textwidth}
\flushleft
\begin{multicols}{2}
\textbf{Research in context}\\
\vspace{0.5em}
\textbf{Evidence before this study}\\
During the development of our artificial intelligence (AI) system, we did a systematic review of the scientific literature with search via PubMed for research articles published on seizure detection with the following inclusion criteria: (1) Tests or inference evaluation is conducted on large-scale clinical EEG data; (2) Generalization is attempted or potentials for generalization is considered, e.g., in commercialized tools; (3) Seizure detection delay and real-time (aka. online) operation were not considered critical in this context as long as the test was conducted on raw EEG data.
Note that ICU seizure detection or portable seizure alert systems are relying on detection delay and real-time needs. Our keywords include {\lq\lq}prospective seizure detection{\rq\rq}, {\lq\lq}automated seizure detection{\rq\rq}, {\lq\lq}non-patient specific seizure detection{\rq\rq}, {\lq\lq}seizure detection on continuous EEG{\rq\rq}, and {\lq\lq}deep learning-based seizure detection{\rq\rq} and {\lq\lq}machine learning-based seizure detection{\rq\rq}. We found that the only two categories of works meet our criteria: two research papers published in 2020 and works published by commercial tools developers. We cited a recent review of 89 deep learning-based seizure detection, all of which are retrospective. One work from Stanford reported seizure detection on all ages (pediatric to adult ages) using post-acquisition EEG recordings and provided an avenue for independent evaluation by providing a test on a publicly available Temple University Hospital (TUH) EEG dataset. The other work pivoted on algorithmic-assisted real-time seizure risk monitoring in continuous EEG in neonatal intensive care unit (NICU) with 128 neonates (32 with seizures) showing about 20\% improvement in seizure identification over 130 neonates (38 with seizures) with no algorithmic assistance.

Commercial tools we studied are Encevis (EpiScan), Besa, and Persyst. There is a recent comparative study on these tools on 81 patients. Encevis is reported as the best performing tool, and hence we provided a comparative study with Encevis ver. 1.9.2. Encevis is also the only tool that provided an avenue for comparative study on publicly available EEG data. The Stanford work, published in 2020, confirms many false positives with Persyst 13. We excluded our tests on Persyst 14 as it highly under-performed relative to Encevis. Only Stanford's work provides code availability. We compared our results with Stanford's work outcome and provided pilot test results with the Encevis (EpiScan) tool on the Australian dataset, which shows a considerably lower sensitivity.

\vspace{0.5em}
\textbf{Added value of this study}\\
To the best of our knowledge, the current study is the first continental generalization that demonstrates the potential to achieve an expert human-level seizure recognition rate in a clinical setting and in just a fraction of time. The two datasets used in this study are recorded with different infrastructure, which adds to the independence of inference from hardware types and improves clinical utility. This is particularly important as 80\% of patients with epilepsy live in low and middle-income countries with limited resources, particularly EEG specialists and neurologists.

\vspace{0.5em}
\textbf{Implications of all the available evidence}\\
Our results support the potential benefits of deep learning AI in clinical settings for seizure recognition and its contribution to significant sensitivity over available solutions. Our AI-assisted system achieves more than a ten-fold increase in time efficiency and reports identical performance to human experts for EEG interpretation with access to great neurophysiology support and auxiliary data. Our findings, particularly our tests on an available commercial tool, recommend that the evaluation, test, or inference in AI systems be performed on different datasets, with diverse infrastructures, and on large-scale and realistic sets with long interictal periods.
\end{multicols}
\end{minipage}
}
\end{figure*}

\section{Introduction}
The lack of continental multi-dataset generalization, non-patient-specificity, prospectiveness, transparency and reproducibility are constantly reported as key challenges toward the broad clinical utility of artificial intelligence (AI) and AI-assisted systems across a broad range of medical data modalities such as imaging, electronic health records, or time-series signals~\cite{haibe2020transparency,kelly2019key,mckinney2020international}. Evaluation of the potential clinical performance of an algorithm is most likely to be achieved in a prospective clinical setting. These shortages in electroencephalography (EEG) data analysis for seizure recognition are apparent in recent reports~\cite{baumgartner2018seizure,koren2021systematic}. For instance, in a recent comprehensive review paper~\cite{shoeibi2020epileptic}, all of the 89 works are retrospective, meaning that they require fully labeled data and lack generalization across multiple large datasets recorded with different infrastructure and hardware. In this paper, we report the first model that provides a combination of continental generalization, non-patient-specificity, inference-only, transparency, and reproducibility and achieves a balance between sensitivity and average rate of false alarms. 

Nearly 30\% of epilepsy diagnoses will not respond to medication, and about 80\% of patients live in low and middle-income countries, with already depleted resources against demands~\cite{baumgartner2018seizure}. The direct and indirect economic burden of epilepsy is high, and automation of seizure detection and labeling could help relax the pressure on resources in epilepsy services clinics.  This research is designed and implemented to benefit epilepsy services clinics by reducing the resources required for labeling recorded EEG data.

\begin{figure*}[ht!]
\centering
    \includegraphics[width=0.95\textwidth]{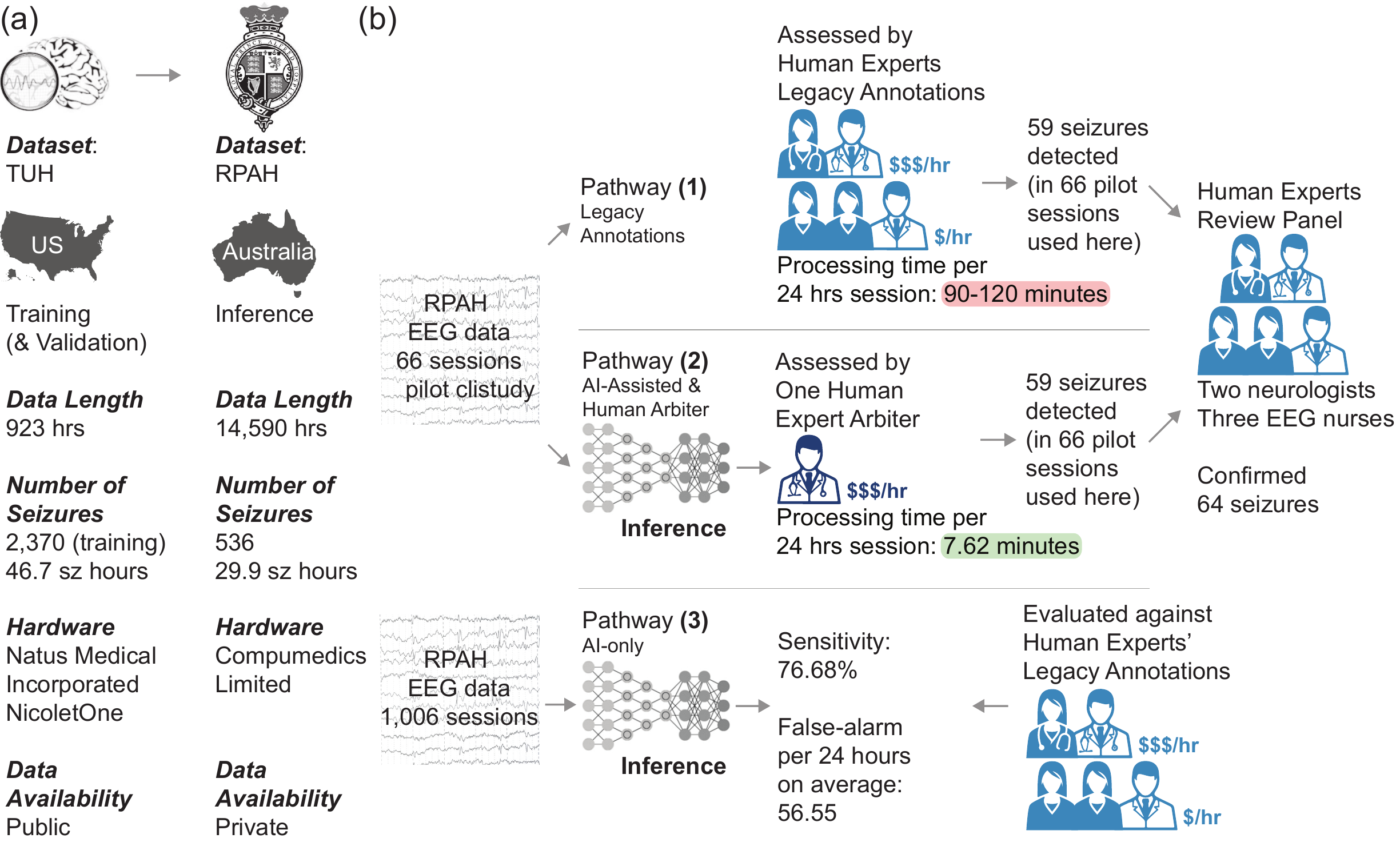}
    \caption{(a) Illustrates the two datasets, across two continents, their hardware difference and also their use in this study. (b) Seizure (Sz) detection and labelling via a conventional pathway (1) and our proposed alternative (2). Pathway (1) represents a commonly used approach in which trained nurses (\$/hour), EEG technicians (\$/hour), and neurologists (\$\$\$/hour) review and label EEG data. In this method, on average, each $24$ hours recording session requires about $90$ to $120$ minutes of review. This time is a conservative measure, as some cases may require more or multiple reviews. Pathway (2) in contrast, represents our proposed alternative in which an expert human arbiter (e.g. a neurologist) is supplied with the output of our AI-system. The AI-system is designed and implemented to achieve high sensitivity (best in class), while maintaining a low false alarm of less than $57$ per $24$ hours on average. Our graphical user interface (GUI) provides the expert with a way to quickly review and accept or reject each onsets (and offsets) of seizure events. We verified our method in this study using multiple arbiters in the series but the time is calculated based on practical equivalence of using only one arbiter in deployment. (Merge) Shows a gateway that for this study approved the performance of the work. This gateway consists of two neurologists and three EEG nurses or technicians. This is to validate the entire system performance against a ground-truth, that is determined by this team of experts. This performance evaluation on 66 recorded EEG sessions resulted in identical performance by the AI-assisted system, pathway (2), and conventional approach, pathway (1).}
    \label{fig:Methods}
\end{figure*}

We report an inference AI system performed on 1,006 sessions that includes 14,590 hours of adult EEG data. This Australia-based data was collected at the Comprehensive Epilepsy Service at the Royal Prince Alfred Hospital (RPAH) in Sydney using a Compumedics Limited EEG infrastructure. We conducted AI model training and validation on a US-based Temple University Hospital (TUH) EEG corpus recorded with Natus Medical Incorporated NicoletOne EEG system. TUH's EEG data length is 923 hours and is focused on seizures, and provides very short interictal data. Fig.~\ref{fig:Methods}(a) provides some preliminary statistics on both datasets. As Fig.~\ref{fig:Methods}(b) suggests, we conducted a three pathways study. First, we performed an inference-only test on the entire RPAH dataset, shown in Fig.~\ref{fig:Methods}(b) Pathway (3). Relative to a legacy ground-truth, which we were not presented with prior to the test, we report 76.68\% sensitivity and, on average, 56.55 false alarms per 24 hours recorded EEG session. The other two pathways are dedicated to clinical tests on randomly selected 66 sessions from RPAH dataset. These 66 sessions are wholly reviewed by an expert panel of two neurologists and three EEG specialists. This results in an updated ground truth for a more accurate analysis of our pilot clinical tests. In Fig.~\ref{fig:Methods}(b) Pathway (2), the AI-generated alarms are reviewed by a board-certified clinician (known in this work as the expert human arbiter) who used our specially designed user interface (UI) for accurate time recordings. The layout and function of this UI is implemented as familiar as possible for the human experts. This analysis confirmed the AI-assisted test achieves an identical performance with the best outcome that was possible to gain with an expert panel of five (92.19\% seizure detection rate with no false alarm), shown in Fig.~\ref{fig:Methods}(b) Pathway (1). Time resources required per 24 hours recorded EEG could be as high as 120 minutes for human only, while the AI-assisted method recorded just 7.62 minutes on average per 24 hours of data.

A recent study in the US Children's Hospital of Philadelphia and the University of Pennsylvania concluded that to achieve $89$\% identification of electrographic seizures in critically ill children; the decision-maker should be willing to pay more than \$22,648 per $48$ hours~\cite{abend2015much}. Additionally, EEG training to prepare expert labor requires a non-trivial amount of full-time study and dedication over six months to two years. The need for clinical care has spurred the emergence of automated non-patient-specific seizure detection algorithms. Among them, deep learning methods provide more accurate and promising ideas for this problem~\cite{golmohammadi2020deep,saab2020weak}. However, existing techniques and solutions still cannot meet the minimum requirements for clinical usage. The major bottleneck is high false alarm if the sensitivity reaches an acceptable level, set by the clinician, who does not want to miss even one seizure or could relax that requirement a bit. Recent analyses have shown that ensuring model generalization across patient populations with different characteristics remains a challenge, necessitating label curation and model retraining to deploy machine learning models to different demographics~\cite{zech2018variable}. Practically, it is unrealistic for patients in the ICU to train a new model for a specific patient and apply it for several days. Besides, creating a new full labeled dataset requires physician-months or physician-years of labeling time, making repeated re-labeling campaigns a substantial diversion of resources. Therefore, A generalized pre-trained automated seizure detection model across the different hospitals, different gender and age patients with high sensitivity, and a reasonably low false-alarm rate is an urgent unmet need for most epilepsy clinics~\cite{furbass2015prospective}. 

As far as we know, there are a number of commercialized seizure documentation tools such as Persyst (Persyst Development Corporation)~\cite{scheuer2020seizure}, Encevis (EpiScan)~\cite{furbass2015prospective,furbass2018eeg} and Besa~\cite{koren2021systematic}. All these tools can perform prospective studies, given they do not need any training or knowledge of data before test or inference. A study performed by Koren~\textit{et al.}~\cite{koren2021systematic} on the three tools reports a higher performance for Encevis, while other independent studies, such as Khaled~\textit{et al.}~\cite{saab2020weak}, report high false alarm rates for Persyst. Unfortunately, the study reported by Khaled~\textit{et al.}~\cite{saab2020weak} did not perform a test against the so far best-reported seizure detection tool available, Encevis. 

Despite several successful reports~\cite{koren2021systematic,furbass2015prospective,furbass2018eeg} and good control over specificity, as we demonstrate in this work, Encevis severely lacks sufficient sensitivity in independent and continental generalization. This is important because seizures in adults are usually and statistically infrequent, and hence, a low sensitivity poses a challenge towards clinical applicability and benefit of the tool.

In this work, we describe a convolutional long short-term memory (ConvLSTM) network~\cite{xingjian2015convolutional} to identify seizures on EEG with high sensitivity. We then use an output lens method, inspired by the early works by Hartmann~\textit{et al.}~\cite{hartmann2011episcan} to monitor our alarms selectively and before they are announced. This helped us balance sensitivity and specificity that is otherwise very challenging to gain for tests in clinical settings.

\section{Method}
\subsection{Study design}
We developed an AI interface embedding a ConvLSTM network as its core for detecting the onset and offset of epileptic seizures. As a continental generalization study framework, we train the AI interface using a US-based dataset from the Temple University Hospital (TUH)~\cite{shah2018temple} and run inference on an Australia-based dataset from the Royal Prince Alfred Hospital (RPAH) to recognize seizure. The RPAH dataset was kept unseen during the training of the AI.

Our AI is trained to achieve a good sensitivity balanced with high specificity, but as a post-processing technique, we used a real-time deterministic method, which we call lens, inspired by Hartmann~\textit{et al.}~\cite{hartmann2011episcan}, to refine the AI's results of seizure recognition. We initially run inference on all raw RPAH datasets (as recorded), and then the legacy ground truth was revealed to us. RPAH EEG recording infrastructure had a significant infrastructure update in 2011, so our analysis only includes data between 2011 and 2019. 

As part of a pilot clinical study then we provide a portion of randomly selected EEG recordings to an expert human arbiter (EEG specialist) to review and provide a quick Go/No Go decision using our implemented user interface for this purpose and to objectively record time for reviews with or without AI involved, which is shown in Fig.~\ref{fig:Methods} Pathways (2) and (1), respectively. Prior to the result assessment, a team of five experts (three EEG specialists and two neurologists) reviewed the data and provided a ground truth for our clinical analysis. 

\subsection{Datasets}
As illustrated in Fig.~\ref{fig:Methods}(a), there are two datasets used in this work: (1) the TUH EEG Corpus~\cite{shah2018temple} solely for training, reproducibility, and future comparative studies, and (2) the RPAH dataset of adult EEG for our inference and clinical tests. For our result assessment post-inference, we were presented with a legacy ground-truth on all 1,006 session data and a ground-truth established by a panel of experts on randomly selected 66 EEG recording sessions on which we ran our clinical test on. The review involved careful visual inspection of recordings by the EEG experts, as shown in Fig.~\ref{fig:Methods}(b). The TUH dataset is the largest publicly available epilepsy database in the US that contains EEG data. We used 1,185 sessions with 592 patients, from which 202 are patients with seizures in our training set, and 238 sessions with 50 patients (40 patients with seizures) in our development set.

\begin{table} 
\caption{Summary of the TUH EEG dataset}
\centering
\resizebox{0.35\textwidth}{!}{
\begin{tabular}{lll}
\toprule
{\bf Dataset attribute} & {\bf Train} & {\bf Dev}\\

\midrule
Files & $4597$ & $1013$\\

Sessions & $1185$ & $238$ \\

Patients & $592$ & $50$ \\

Files with seizures & $867$ & $280$ \\

Sessions with seizures & $343$ & $104$ \\

Patients with seizures & $202$ & $40$ \\

Number of seizures & $2370$ & $673$ \\

Background duration (hours) & $705.6$ & $154.1$ \\

Seizure duration (hours) & $46.7$ & $16.2$ \\
\midrule
Total duration (hours) & $752.3$ & $170.3$ \\

\bottomrule
\end{tabular}
}
\label{tab:data}
\end{table}

\begin{figure*}[ht!]
\centerline{
    \includegraphics[width=0.90\textwidth]{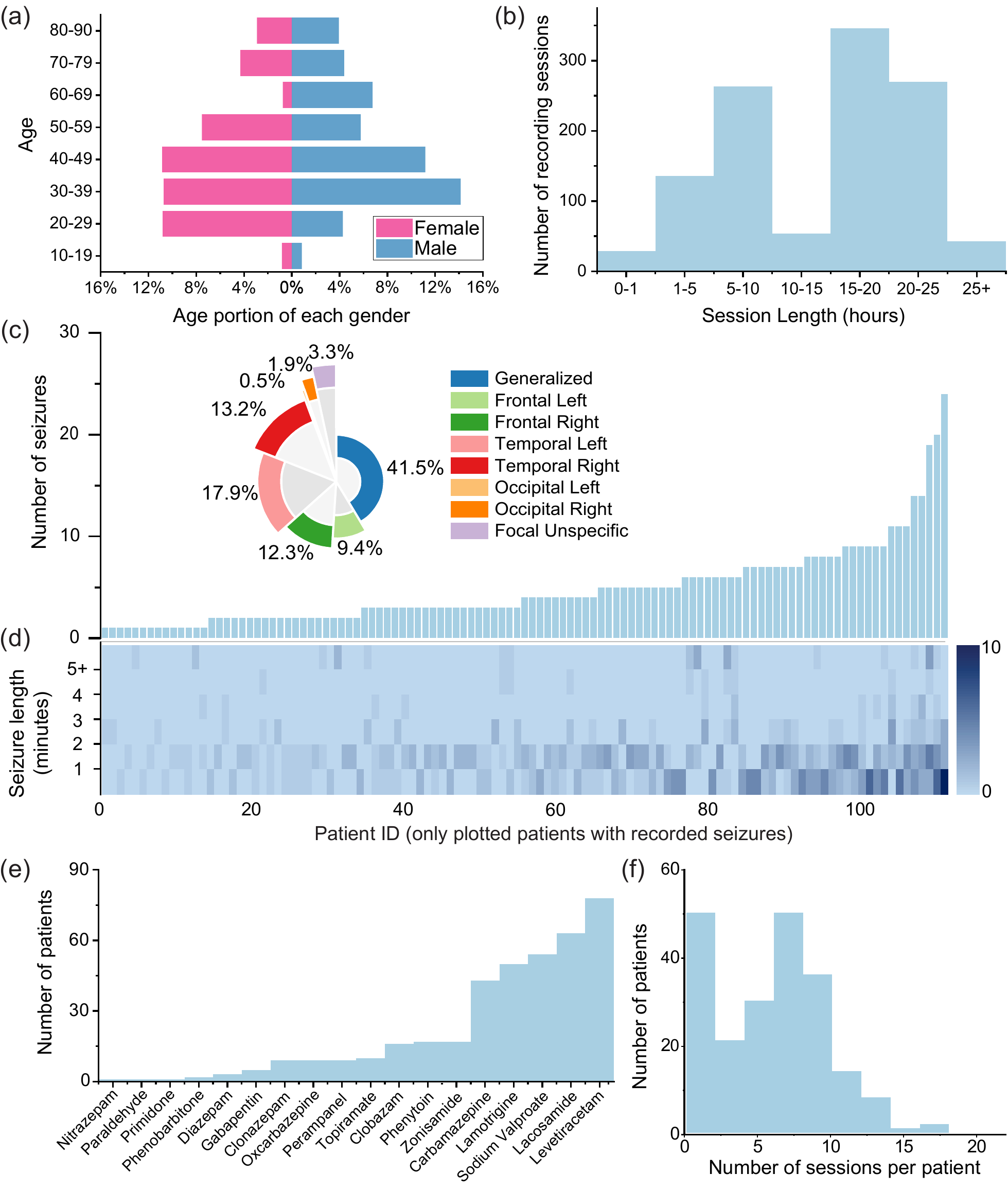}}
    \vskip -.3cm
    \caption{Summary of the Royal Prince Alfred Hospital (RPAH) dataset. {(a)} Patients' age distribution and gender (inset). {(b)} Monitoring sessions lengths. {(c)} Distribution of the number of seizures per patient (only those with detected and documented seizures are plotted, based on the final ground-truth), and their seizure types (inset). {(d)} Heat-map of seizure lengths for each patient with detected seizures; Changing color from shallow blue to dark blue represents an increase in the number of seizures in that band. {(e)} Histogram of anti-epileptic drugs (AED) administered for patients (AED types may overlap in a given patient). {(f)} The number of EEG recording files per patient.}
    \label{fig:RPA_dataset}
\vskip -.2cm
\end{figure*}

To verify the proposed system's clinical utility, we test the trained model with an inference-only mode on the RPAH EEG dataset. An ethics approval was acquired to support our access to this clinical data. RPA Hospital is one of Australia's major hospitals, with one of the longest, if not the longest, EEG recordings in Australia in its Comprehensive Epilepsy Services. RPA structurally and reliably maintained data of its adult epilepsy patients, who came from across Australia. In this work, we select nine years (2011--2019) data to test with nearly 14,590 hours of EEG data, from 192 patients with a total of 1,006 sessions, each of which has an average recording length of around 15 hours. The number of patients in the data is 212, which we excluded 20 for different reasons, including too many seizures (measured in more than 11 seizures/~24 hours) (5 patients), missing electrode data (14 patients) or seizures can only be confirmed by video information (1 patient). The detailed information is shown in Fig.~\ref{fig:RPA_dataset} for RPAH dataset. The neurologists randomly select the 66 clinical test sessions without any prior information of the patient's history.

\subsection{Pre-processing}
Although raw EEG data information can be directly fed into a neural network, the lack of frequency information mixed with artifacts will make it harder for the network to extract essential features~\cite{fisher1992high}. To address this, we used two signal processing techniques, independent component analysis (ICA)~\cite{comon1994independent} and short-time Fourier transform (STFT). 

First, we split EEG signals into 12-second segments and applied the ICA algorithm to decompose the signal into $19$ independent components using the Blind Source Separation (BSS) approach~\cite{belouchrani1997blind}. The Eq.~\ref{ICA} shows the principle of the BSS. The ICA algorithm assumes that the matrix $A$ contains statistically independent topographic maps, and $M$ represents the time courses. 
\begin{align}
     \label{ICA}
   T\approx MA^\top,
\end{align}
where $T\in\mathbb{R}^{I_t\times I_e}$ represents multiple channel EEG signals; $I_t$, and $I_e$ represent the number of samples in time and number of electrodes, respectively. After decomposition, $M\in\mathbb{R}^{I_t\times R}$ has the time information and 
$A\in\mathbb{R}^{I_e\times R}$ contains the topographic maps weights, where $R$ is the estimated number of independent sources.

We use Pearson correlation to identify which independent sources are highly related to eye movement that is detected from two EEG channels, namely {\lq}FP1{\rq} and {\lq}FP2{\rq}~\cite{dammers2008integration}. We remove those independent sources and reconstruct the EEG signals to obtain the eye movement artifact-free signals. Hence, we perform STFT on clean EEG signals using a window length of 250 (or 1 second) and 50\% overlapping, then remove the DC component of the transform so that the data shape will become ($n\times23\times125$), where $n, 23$ and $125$ represent the number of electrodes, the time index, and the frequencies, respectively. Our artifact removal is implemented in Python 3.6 with the use of library MNE v0.20.~\cite{gramfort2013meg}.

\begin{figure*}[ht!]
\centering
    \includegraphics[width=1.0\textwidth]{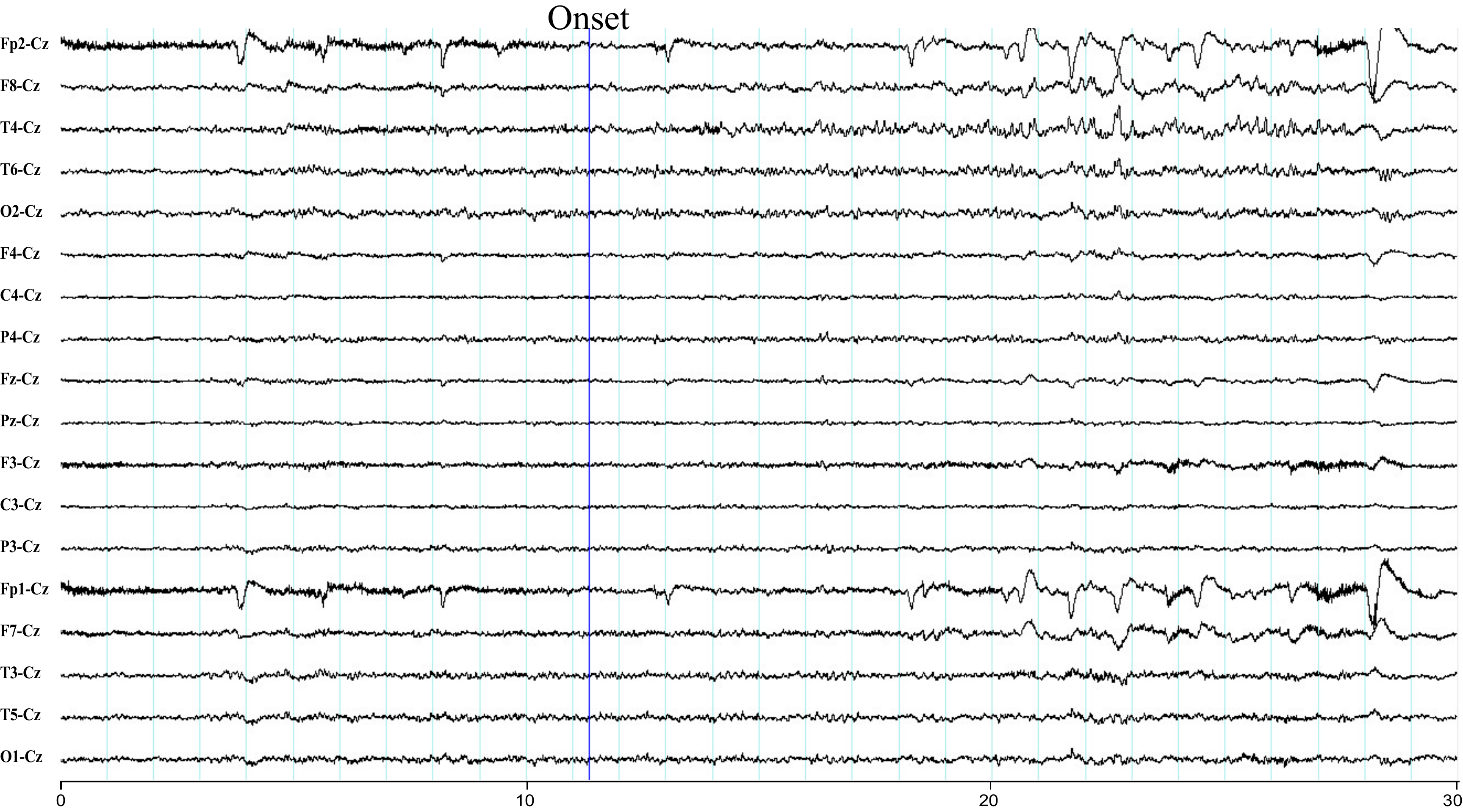}
    \caption{Example of extra seizure detected by AI. The human arbiter missed this seizure at first but AI detected it. A review by a panel of two neurologists and three EEG specialists agreed that it was a valid seizure detection. It can be seen that rhythmic sharp alpha evolving to theta in the right temporal region (T4 and T6). More examples that human missed but AI detected are shown in the Supplementary information Figs.~2,3,4,5.}
    \label{fig:extra_seizure}
\vskip -.5cm
\end{figure*}

\subsection{Machine learning interface}
\begin{figure*}[ht!]
\centering
    \includegraphics[width=1.0\textwidth]{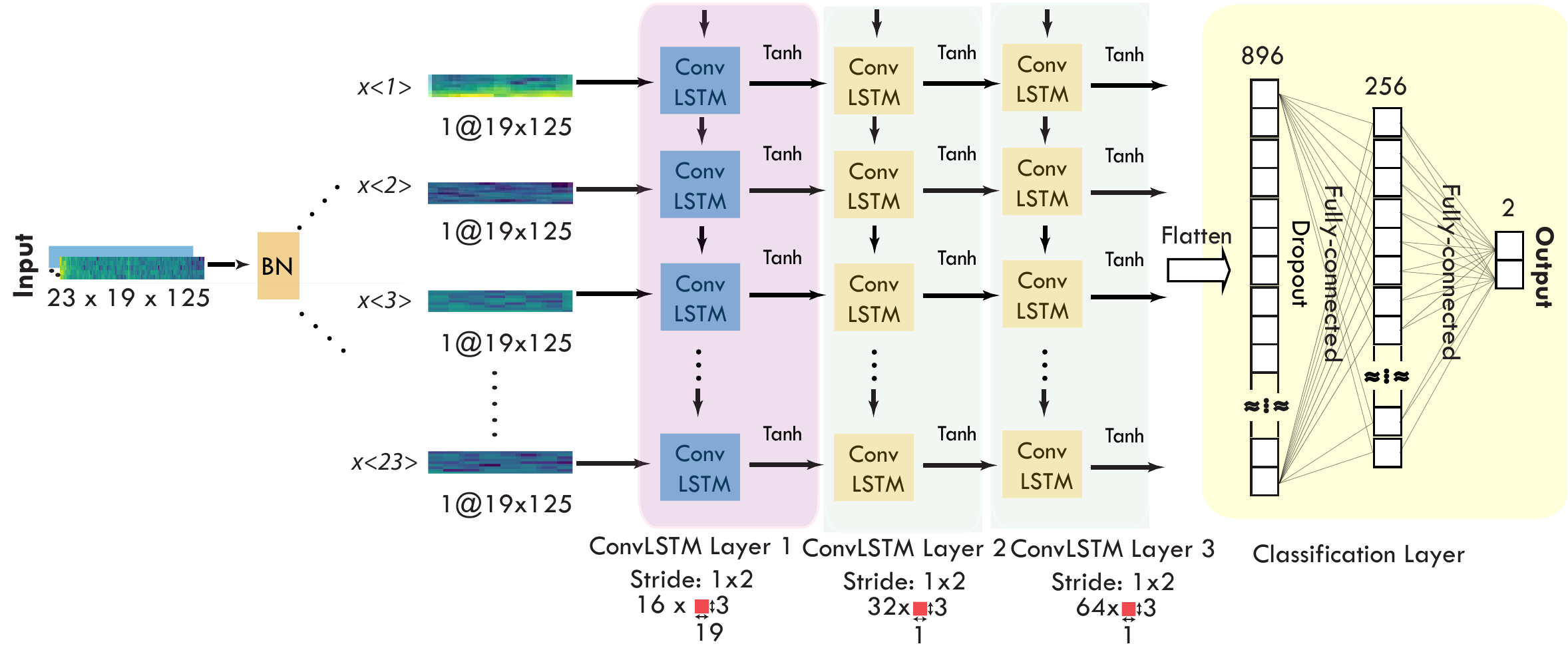}
    \caption{Deep Learning Model}
    \label{fig:deep_model}
\vskip -.2cm
\end{figure*}

We built a machine learning interface that displays the raw EEG signals and the corresponding probability of seizure suggested by AI, which helps the clinician annotate seizures. The interface also records the review time for each session automatically. (one example is shown in the Supplementary information Fig.~1). At the core of the interface is a deep learning model consisting of three ConvLSTM blocks~\cite{xingjian2015convolutional}, followed by three fully connected layers. The detailed structure is shown in Fig~\ref{fig:deep_model}. The first ConvLSTM layer has 16 ($n \times 19 \times 3$) kernels using ($1 \times 2$) stride, where {\it n} represents the number of channels. The next two ConvLSTM blocks both use ($1 \times 2$) stride and ($1 \times 3$) kernel size. The number of kernels is 32 and 64 for the ConvLSTM blocks 2 and 3, respectively. Two fully connected layers follow the three ConvLSTM blocks with sigmoid activation and output sizes of 256 and 2.

Our model is implemented in Python 3.6 with the use of Keras 2.0 and Tensorflow 1.4.0. We use a dropout layer with 0.5 probability applied into all fully connected layers to avoid over-fitting. We also apply the early-stopping technique to stop the training when the combined training and valuation loss have not decreased for 20 consecutive epochs. Besides, the model training is accomplished using Adam optimizer, and the learning rate is set as $5\times10^{-4}$. Training the model with the TUH train set took approximately one day with an NVIDIA V100 GPU.

\subsection{Real-time post-processing}
After the machine learning model is trained with the TUH dataset, we run it in the inference mode directly on the RPAH without any further training or knowledge of RPAH annotations. In this process, the high risk (probability $\geq 10\%$) seizure areas detected by the AI are sent to a lens focus algorithm in charge of reviewing the results and generating alerts.

The lens is a real-time signal processing method called periodic waveform analysis (PWA), initially presented by Hartmann~\textit{et al.}~\cite{hartmann2011episcan}. Periodic energy index (PEI) and periodic waveform index (PWI) values for alpha, beta, theta, delta, gamma rhythms were calculated and, through an automated and adaptive process of threshold setting for the power in each band (rhythm), the lens can identify most likely ictal events. 

First, the total harmonic energy ($E_{\tau}$) is calculated within a certain period ($t_{min}\leq\tau\leq t_{max}$) of EEG signal ($x_{t}$), where $\psi^{*}$ is the time window to extract the signal information. The PEI value is defined as the maximum value in that periodical area, which is shown in the equation below:   
\begin{equation}
    E_{\tau} = \sum_{m>0}\Bigg|\frac{1}{\sqrt{\tau}}\int_{-\infty}^{\infty}x_{t}\psi^{*}_{\frac{t}{\tau}}e^{-j2\pi\frac{mt}{\tau}} dt \Bigg|^{2}
\end{equation}
\begin{equation}
PEI = \max_{t_{min}\leq\tau\leq t_{max}}E_{\tau}
\end{equation}
Then the signal energy value is calculated in that period. 
\begin{equation}
N_{\tau} = \frac{1}{\sqrt{\tau}}\int_{-\infty}^{\infty}\big|x_{t}\psi^{*}_{\frac{t}{\tau}}\big|^{2}dt
\end{equation}
The PWI value is defined as:
\begin{equation}
PWI = \frac{ E_{\tau}}{N_{\tau}}
\end{equation}
Then, the $PWI_{\alpha}$, $PWI_{\beta}$, $PWI_{\theta}$, $PWI_{\delta}$, $PWI_{\gamma}$ values are split based on different brain signal frequency bands (0--3~Hz, 4--7~Hz, 8--12~Hz, 13--30~Hz, and more than 31~Hz). The same value split is applied for the PEI. We use the 85-percentile of PWI and PEI values for each frequency band over the last two hours as adaptive thresholds. If the PWI and PEI values are higher than the corresponding adaptive thresholds in all frequency bands, the period will be reported to the interface. Using this technique, we achieved a significantly higher sensitivity than EpiScan techniques while maintaining an acceptable level of false alarms when tested with RPA Hospital data.

\subsection{Clinical test with human arbiters}
After the post-processing, the potential seizure areas are highlighted in the interface. Our human arbiter committee consists of two board-certified practicing epilepsy neurologists and three board-certified practicing EEG specialists. All committee members only review the high-risk area highlighted by the machine learning interface and decide based on their clinical experience. The final results are determined by the panel's majority votes and then compared with the ground-truth results.

\subsection{Performance metrics}
\label{metrics}
To assess how effectively the proposed method performs for the seizure detection task, we compute the Area Under the Receiver Operating Characteristic curve (AUC), sensitivity or true positive rate, false-positive rate (FPR), seizure detection rate (SDR), and false alarm per 24 hours (FA/24 hours).

The SDR rate is defined as the number of seizures detected over the total number of seizures. The AUROC score is used to measure the ability of the machine learning model to classify the seizure and non-seizure clips regardless of the threshold selection. The values of recall and FPR can be derived from the ROC curve. In the clinical test, the seizure count and the number of false alarms are two critical indicators of the method's usefulness to the patients and clinicians. Therefore, we use SDR and FA/24 hours as our main metrics in this study.

\section{Results}
\begin{figure}[ht!]
\centering
    \includegraphics[width=0.5\textwidth]{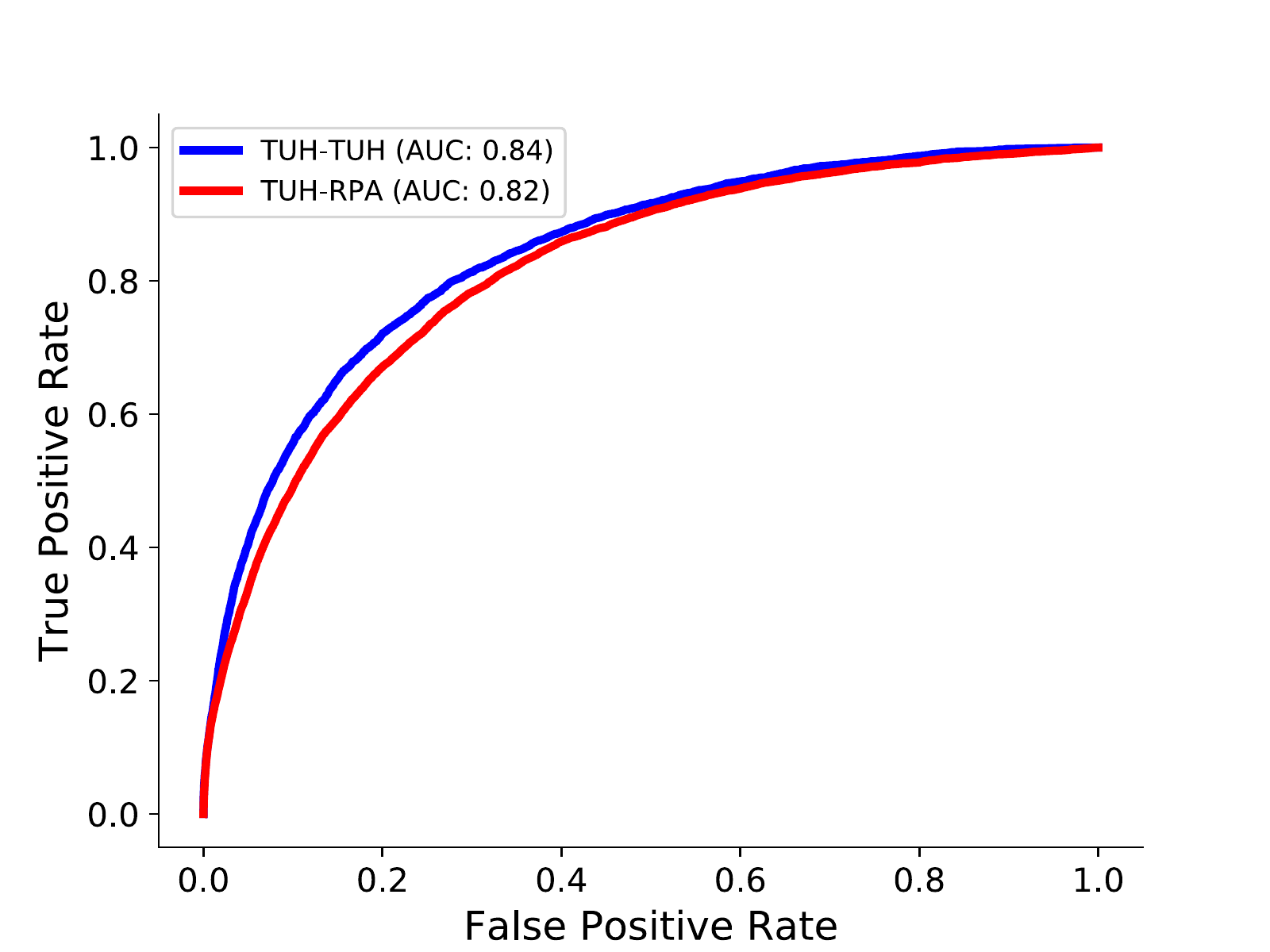}
    \caption{Receiver operating characteristic (ROC) curves. Two curves using the area under the ROC curve (AUC) metric. TUH-TUH represented the model being trained on the TUH training dataset and tested on the TUH development dataset. TUH-RPAH represented the model being trained on the TUH training dataset and tested on the RPAH dataset.}
    \label{fig:RPA_ROC}

\end{figure}

We trained the ConvLSTM model on the TUH dataset and performed the inference on the RPAH dataset. The overall results are shown in Table~\ref{tab:result}. Our results on the TUH dataset provide an avenue of comparative assessment with similar works such as the one proposed by Khaled~\textit{et al.}~\cite{saab2020weak}. Relative to that work, we improved the average AUC score by $6\%$ (see Fig.~\ref{fig:RPA_ROC}) while using $12$-second moving time windows in our analysis compared with $60$-second in theirs. A longer time window generally results in an improved AUC score~\cite{saab2020weak}. Some highly regarded works that were trained, validated, and tested on the TUH dataset are reported in Table~\ref{tab:result}. These retrospective studies are likely to lack clinical utility, as they are limited by design~\cite{shah2017optimizing,golmohammadi2020deep}, and they achieved a low sensitivity score using OVLP method~\footnote{OVLP refers to {\lq\lq}Any Overlap Metric{\rq\rq}~\cite{ziyabari2017objective}.}. It is expected that the short interictal periods in the TUH dataset do not provide a realistic specificity test venue for any seizure detection research and development, unfortunately. The TUH training set contains 46.7 hours of seizures, which is 6.2\% of the dataset, while the RPAH dataset includes 29.9 hours of seizures, which is 0.2\% of the aggregated dataset duration. This shortage, alongside retrospective models, is appeared in IBM's model-to-data privacy-oriented platform for seizure documentation on the TUH dataset~\cite{roy2021evaluation}.

An inference of the method on 14,590 hours of RPAH set achieved an AUC of 0.82 (Fig.~\ref{fig:RPA_ROC}), 76.68\% sensitivity, and on average 56.22 false alarms per 24 hours via the commonly used metric of SDR (results in detail are provided in the Supplementary information Table~II). To explore the efficiency of the proposed AI-assisted system in the clinical setting, we did a clinical test on $66$ sessions randomly selected by neurologists. Our proposed method (Fig.~\ref{fig:Methods}(b) Pathway (2)) achieved 92.19\% sensitivity, in line with the conventional method in Fig.~\ref{fig:Methods}(b) Pathway (1), but results in a time and financial saving of up to 10$\times$ that of the conventional method. A prospective, multi-center study by Furbass~\textit{et al.}~\cite{furbass2015prospective} achieved promising results using conventional signal processing methods. Their work only detects seizure onsets and presents a significant disadvantage when it comes to independent sensitivity tests on large-scale datasets, as we demonstrated in our work. We tested their tool, Encevis (EpiScan), in our clinical test, achieving a relatively low sensitivity of 62.50\% (compared with the proposed method 92.19\%). Encevis achieved a low false alarm number of 7.02 per 24 hours. In the clinical environment, sensitivity is most important for neurologists. Thus, our AI system balances both sensitivity and false alarms in our proposed framework by having the human arbiter reject the false alarm. 

In the 66 sessions clinical test, our model missed eight (8) seizures (after reviewing by neurologists, only 5 of them are confirmed as valid seizures) compared with seizure detection done by clinicians. By combining our proposed model and the expert arbiter, the human arbiter observed five additional seizures in comparison to the first review (diagnosis of epilepsy routine). The three factors used to compare our proposed method with the conventional human method are accuracy, time, and financial cost. Our model is matched on the accuracy, but superior in terms of time, hence, financial cost. A significant reduction in time for reviewing EEG signals was observed for each 24 hours recording (7.62 minutes versus 90 minutes), as shown in Fig.~\ref{fig:RPA_time} and in detail in the Supplementary information Table~I).

\begin{table*}[ht!]
    \centering
    \caption{Results comparison}
    \label{tab:result}
    \resizebox{\textwidth}{!}
    {        
        \begin{tabular}{ c*{11}{l}  }    
            \toprule

\multirow{2}{2.21cm}{Dataset} & 
\multirow{2}{1.55cm}{Prospective~$^\dagger$} &
\multirow{2}{1.75cm}{International\\ Generalization} &
\multirow{2}{1.55cm}{Reprodu-\\cibility~$^\eta$} &
\multirow{2}{4.55cm}{Method} &
\multirow{2}{2.66cm}{Reference} & 
\multirow{2}{1.55cm}{Seizure length} &
\multirow{2}{0.80cm}{AUC} & 
\multirow{2}{1.55cm}{Evaluation\\method} & 
\multirow{2}{1.35cm}{Sensitivity} &
\multirow{2}{1.80cm}{FA/$24$ hours\\(average)} 
\\ \\
            
            \midrule
            \midrule
            
\multirow{2}{2.21cm}{NCR, MUV, KEMP} & 
\multirow{2}{1.55cm}{Y} &
\multirow{2}{1.75cm}{Y} &
\multirow{2}{1.55cm}{N} &
\multirow{2}{4.55cm}{Deterministic signal processing} &
\multirow{2}{2.66cm}{EpiScan~\cite{furbass2015prospective}} & 
\multirow{2}{1.55cm}{N~$^\xi$} &
\multirow{2}{0.80cm}{$-$} & 
\multirow{2}{1.55cm}{SDR$^\ast$} & 
\multirow{2}{1.35cm}{72.00\%} &
\multirow{2}{1.55cm}{7.05} 
\\ \\

            \midrule

\multirow{2}{2.21cm}{TUH EEG Corpus v1.1.0} &
\multirow{2}{1.55cm}{N} &
\multirow{2}{1.75cm}{N} &
\multirow{2}{1.55cm}{N} &
\multirow{2}{4.55cm}{AI} &
\multirow{2}{2.66cm}{Golmohammadi \textit{et al.} \cite{shah2017optimizing}} &
\multirow{2}{1.55cm}{Y} &
\multirow{2}{0.80cm}{$-$} &
\multirow{2}{1.55cm}{OVLP} &
\multirow{2}{1.35cm}{39.15\%} &
\multirow{2}{1.55cm}{22.83} &
\\ \\

            \midrule

\multirow{2}{2.21cm}{TUH EEG Corpus v1.4.1} &
\multirow{2}{1.55cm}{N} &
\multirow{2}{1.75cm}{N} &
\multirow{2}{1.55cm}{N} &
\multirow{2}{4.55cm}{AI} &
\multirow{2}{2.66cm}{Golmohammadi~\textit{et al.}~\cite{golmohammadi2020deep}} &
\multirow{2}{1.55cm}{Y} &
\multirow{2}{0.80cm}{$-$} &
\multirow{2}{1.55cm}{OVLP} &
\multirow{2}{1.35cm}{30.83\%} &
\multirow{2}{1.55cm}{6.75} &
\\ \\

            \midrule
            \midrule
            
\multirow{2}{2.21cm}{TUH EEG Corpus v1.4.1} &
\multirow{2}{1.55cm}{N} &
\multirow{2}{1.75cm}{N} &
\multirow{2}{1.55cm}{Y} &
\multirow{2}{4.55cm}{AI} &
\multirow{2}{2.66cm}{Khaled~\textit{et al.}~\cite{saab2020weak}} &
\multirow{2}{1.55cm}{Y} &
\multirow{2}{0.80cm}{0.78} &
\multirow{2}{1.55cm}{$-$} &
\multirow{2}{1.35cm}{$-$} &
\multirow{2}{1.55cm}{$-$} &
\\ \\

            \midrule
            
\multirow{2}{2.21cm}{Stanford Hospital} &
\multirow{2}{1.55cm}{Y} &
\multirow{2}{1.75cm}{N~$^\ddagger$} &
\multirow{2}{1.55cm}{Y} &
\multirow{2}{4.55cm}{AI} &
\multirow{2}{2.66cm}{Khaled~\textit{et al.}~\cite{saab2020weak}} &
\multirow{2}{1.55cm}{Y} &
\multirow{2}{0.80cm}{0.70} &
\multirow{2}{1.55cm}{$-$} &
\multirow{2}{1.35cm}{$-$} &
\multirow{2}{1.55cm}{$-$} &
\\ \\

            \midrule
            \midrule
            
\multirow{2}{2.21cm}{TUH EEG Corpus v1.5.1} &
\multirow{2}{1.55cm}{N} &
\multirow{2}{1.75cm}{N} &
\multirow{2}{1.55cm}{Y} &
\multirow{2}{4.55cm}{AI} &
\multirow{2}{2.66cm}{This work} &
\multirow{2}{1.55cm}{Y} &
\multirow{2}{0.80cm}{0.84} &
\multirow{2}{1.55cm}{$-$} &
\multirow{2}{1.35cm}{$-$} &
\multirow{2}{1.55cm}{$-$} &
\\ \\

            \midrule
            
\multirow{2}{2.21cm}{RPAH (1,006 sessions)} &
\multirow{2}{1.55cm}{Y} &
\multirow{2}{1.75cm}{Y} &
\multirow{2}{1.55cm}{Y} &
\multirow{2}{4.55cm}{AI inference,\\Fig.~\ref{fig:Methods}(b) Pathway (3)} &
\multirow{2}{2.66cm}{This work} &
\multirow{2}{1.55cm}{Y} &
\multirow{2}{0.80cm}{0.82} &
\multirow{2}{1.55cm}{SDR} &
\multirow{2}{1.35cm}{76.68\%} &
\multirow{2}{1.55cm}{56.55} &
\\ \\

            \midrule

\multirow{2}{2.21cm}{RPAH (66 sessions pilot)} &
\multirow{2}{1.55cm}{Y} &
\multirow{2}{1.75cm}{Y} &
\multirow{2}{1.55cm}{Y} &
\multirow{2}{4.55cm}{AI inference + Human arbiter, Fig. \ref{fig:Methods}(b), Pathway (2)} &
\multirow{2}{2.66cm}{This work} &
\multirow{2}{1.55cm}{Y} &
\multirow{2}{0.80cm}{$-$} &
\multirow{2}{1.55cm}{SDR} &
\multirow{2}{1.35cm}{92.19\%} &
\multirow{2}{1.55cm}{0} &
\\ \\ 

            \midrule
            \midrule

\multirow{3}{2.21cm}{RPAH (66 sessions pilot)} &
\multirow{3}{1.55cm}{$-$} &
\multirow{3}{1.75cm}{$-$} &
\multirow{3}{1.55cm}{$-$} &
\multirow{3}{4.55cm}{Five human\\experts performance,\\Fig.~\ref{fig:Methods}(b) Pathway (1)} &
\multirow{3}{2.66cm}{This work} &
\multirow{3}{1.55cm}{$-$} &
\multirow{3}{0.80cm}{$-$} &
\multirow{3}{1.55cm}{SDR} &
\multirow{3}{1.35cm}{92.19\%} &
\multirow{3}{1.55cm}{0} &
\\ \\ \\

            \midrule

\multirow{3}{2.21cm}{RPAH (66 sessions pilot)} &
\multirow{3}{1.55cm}{$-$} &
\multirow{3}{1.75cm}{$-$} &
\multirow{3}{1.55cm}{$-$} &
\multirow{3}{4.55cm}{Conventional signal processing} &
\multirow{3}{2.66cm}{Our test on \\Encevis (EpiScan)~\cite{furbass2015prospective}} &
\multirow{3}{1.55cm}{N~$^\xi$} &
\multirow{3}{0.80cm}{$-$} &
\multirow{3}{1.55cm}{SDR} &
\multirow{3}{1.35cm}{62.50\%} &
\multirow{3}{1.55cm}{7.02} &
\\ \\ \\  

            \bottomrule            
        \end{tabular}    
}
\begin{tablenotes}
      \scriptsize
       \item Note: The metrics AUC, SDR (seizure detection rate), FA (false-alarm)/24 hours are detailed in Section~\ref{metrics}. OVLP refers to {\lq\lq}Any Overlap Metric{\rq\rq}~\cite{ziyabari2017objective}. The evaluation method OVLP considers the result correct if the detection is within the reference event or multiple shorter events detected within the long reference event. The sensitivity, FA/24 hours, refers to the sensitivity and the number of false alarms per 24 hours and is calculated by the corresponding evaluation method. In clinical settings, neurologists place more importance on seizure timing and frequency. Thus, seizure detection rate (SDR) is suitable for real-world applications.
       \item $\dagger$ A large cohort of retrospective studies was not included in this paper~\cite{shoeibi2020epileptic}.
       \item $\ddagger$ Khaled~\textit{et al.}~\cite{saab2020weak} generalised across two US-based datasets. There is no information on the hardware of the target dataset.
       \item $\eta$ our initial tests on Persyst, consistent with reports such as Khaled~\textit{et al.}~\cite{saab2020weak}, were not promising. We found reproducibility of results, such as those reported by Scheuer~\textit{et al.}~\cite{scheuer2020seizure}, on other datasets is challenging.
       \item $\xi$ limited to the detection of seizure onset only.
       \item $\ast$ SDR method combines the false alarms within $30$ seconds into one.       
       \item NCR: Neurological Center Rosenhuegel in Vienna, Austria
       \item MUV: Medical University of Vienna, Austria 
       \item KEMP: Epilepsy Center Kempenhaeghe in Heeze, the Netherlands
       \item TUH: Temple University Hospital, USA
       \item RPAH: Royal Prince Alfred Hospital, Australia
\end{tablenotes}

\end{table*}

\begin{figure}[ht!]
\centering
    \includegraphics[width=0.5\textwidth]{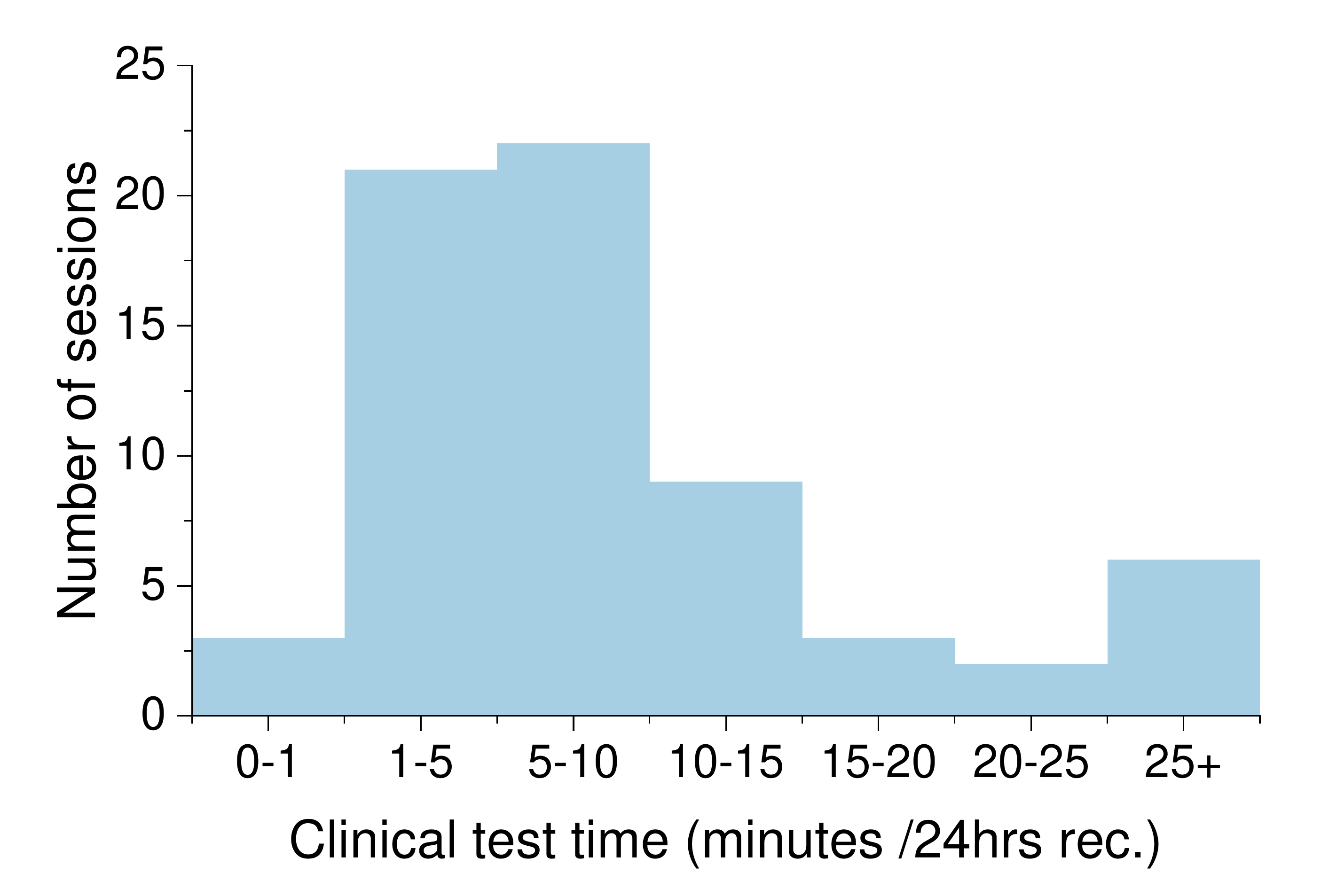}
    \caption{Time consumption of the 66 sessions during the clinical test. This is a histogram showing the actual human time spent on each clinical session. The average time cost is, on average, 7.62 minutes per 24 hours recording---the detail in Supplementary Information Table I.}
    \label{fig:RPA_time}
\vskip -.2cm
\end{figure}

\section{Discussion}

We conducted a continentally generalized inference-only test study on a large-scale Australian EEG dataset (from RPAH) to assess the clinical utility of seizure recognition. We initially trained our AI model on a US-based dataset (from TUH). We trained our deep learning model on two types of information: 1) background information, including interictal phases, and 2) seizure information - the {\lq\lq}ictal phase{\rq\rq}, being seizure onset to seizure ends. First, we tested our model on the TUH dataset, achieving a $0.84$ AUROC score (see Table~\ref{tab:result}) using a $12$-s window, surpassing the score achieved by Khaled~\textit{et al.}~\cite{saab2020weak}. Although the AUC score only improves by 0.06, we use a 12-second window, five times shorter than Khaled ~\textit{et al.}. The 12-second window was selected based on the substantial number of seizures that were less than $60$-s. The F1-score in Khaled~\textit{et al.}, achieved through a $12$-s window, were $0.1$ and 0.27 smaller than the 60-second window on the pediatric and adult LPCH (Luckile Packard Children's Hospital) dataset, respectively. Other remotely relevant work includes one presented by Pavel~\textit{et al.}~\cite{pavel2020machine}, that compared an algorithmic-assisted real-time seizure risk monitoring in continuous EEG for neonatal intensive care unit (NICU) with 128 neonates (32 with seizures) showing about 20\% improvement in seizure identification over 130 neonates (38 with seizures) with no algorithmic assistance.

To determine the generalisability of the model, we tested the pre-trained network directly on the RPAH dataset. The dataset includes 1,006 sessions; our AUROC score reaches 0.82, which is slightly lower than testing with the TUH dataset. The detailed comparison is shown in Fig.~\ref{fig:RPA_ROC}. We achieved an average of 76.68\% sensitivity and 56 false alarms per 24 hours. The results bear great promise for clinical application as our dataset includes various participants, some with seizures that are challenging to detect. Of particular interest is the increase in the number of seizures detected by the AI missed by the clinicians' first review. Therefore the addition of AI in the clinical environment can improve seizure detection rates beyond the results of our study.

Furthermore, we did a prospective clinical test on the 66 EEG recording sessions that neurologists randomly selected. We asked two epilepsy neurologists, and three experienced EEG specialists to label the seizure without the AI interface's help. The results are shown in Fig.~\ref{fig:Methods}. We can see that human arbiter can find 59 out of 64 seizures by reviewing all the EEG recordings. In comparison, with AI's help, human arbiter only needs to check the area with high susceptibility, saving lots of time and move cost but still maintain high accuracy. Overall, it takes 1-1.5 hours and 1.5-2 hours on average for a neurologist and an EEG specialist to review a 24 hours surface EEG recording with an approximate 85\% and 75\% accuracy in the first review, respectively.
Interestingly, with AI's help, one human arbiter found 15 potential extra seizures that were not labeled and missed eight seizures that the AI did not highlight. The five human arbiters confirmed 5 out of 15 are valid seizures that were not detected during the first time of review (without AI). One of the five extra seizures is shown in Fig.~\ref{fig:extra_seizure}, and neurologists found that these seizures have a common characteristic: short, subtle frequency evolution. Thus it is hard to identify when human arbiters first time review the whole recordings. For the eight seizures that the AI missed, three of them could only be confirmed by reviewing video recordings. These seizures could not be detected by visual inspection of EEG signals (examples are shown in the Supplementary information Figs.~6,~7,~8). For the other five missed seizures, neurologists found that the majority of them were very brief clinical seizures. Understandably, short seizures are quite hard to detect as the EEG biomarker or patterns could be ambiguous. Another weakness of AI is the ability to detect clinical seizures. For this type of seizure, the patients usually report the events despite no change in the surface EEG signals. It is known that not all seizures are associated with surface EEG change~\cite{goetz2007textbook}. Overall, with the help of AI, the human arbiter finds 59 out of 64 seizures. The result shows that the proposed method saves around ten times in terms of money and time cost, while the error rate remains unchanged.

\section{Conclusion}
Seizure prediction and detection capability have been studied and improved over the last four decades. A board-certified EEG specialist is required by law to diagnose epilepsy. However, it takes several years to train a clinician, and the ability to generate data far exceeds the human ability to translate the data.
Therefore, a reliable generalized AI-assisted detection system will relieve the clinician's work and help a patient have a more manageable life. In the meantime, the false alarm is critical to the application as it impacts clinicians and patients' workload. Our proposed method shows the advantage of largely reducing the time and money cost while maintaining a high accuracy level and can apply directly into clinical without acquiring the training data.

\section*{Acknowledgements}
The authors would like to thank neurologists Armin Nikpour, Kaitlyn Sharp, and EEG specialists Sumika Ochida, Maricar Senturias, Toh Hock Wong, IT technologist Satendra Pratap at the Comprehensive Epilepsy Service at the Royal Prince Alfred Hospital (RPAH), Sydney, Australia, for their dedication and contribution. The authors would also like to thank Sheng-Ku Lin for his logistical help during the time-consuming process of converting the EEG data. The authors would also like to thank Dr. Jason Eshraghian for his insightful inputs on our results. Yikai Yang would like to acknowledge the Research Training Program (RTP) support provided by the Australia Government. Omid Kavehei acknowledges the support provided by The University of Sydney through a SOAR Fellowship and Microsoft's support through a Microsoft AI for Accessibility grant.

\section*{Ethics declarations}
Ethics X19-0323-2019/STE16040: Validating epileptic seizure detection, prediction, and classification algorithms approved on 19 September 2019 by NSW Local Health District (LHD) for implementation within the Comprehensive Epilepsy Services at the Department of Neurology at the Royal Prince Alfred Hospital.

\section*{Data sharing}
The Temple University Hospital dataset is publicly available at \url{https://www.isip.piconepress.com/projects/tuh_eeg/html/downloads.shtml}. The Department of Neurology dataset at the Royal Prince Alfred Hospital was used under ethics Review Board approval for use only in the current study, which is not publicly available yet, but we hope to make these data available in the future, pending proper approval through the ethics committee.

\section*{Declaration of interests}
The authors declare no competing interests.

\section*{Authors' contributions}
YYK contributed to the data curation, formal analysis, investigation, methodology, resources, software, visualization, validation, writing original draft, writing review \& editing. NNT contributed to data curation, formal analysis, investigation, methodology, resources, software, supervision, visualization, validation, writing review \& editing. CM contributed to data curation, resources, validation, writing review \& editing. AN contributed to conceptualization, data curation, funding acquisition, investigation, project administration, resources, supervision, validation, writing review \& editing. OK contributed to conceptualization, data curation, funding acquisition, investigation, methodology, project administration, resources, supervision, visualization, validation, writing review \& editing.

\section*{Code availability}
The code used to generate all results in this manuscript can be made available upon request.

\newpage
\section{Supplementary Information}
\subsection{Method}
\subsubsection{Machine Learning Interface}
\begin{figure*}[ht!]
\centering
    \includegraphics[width=1.0\textwidth]{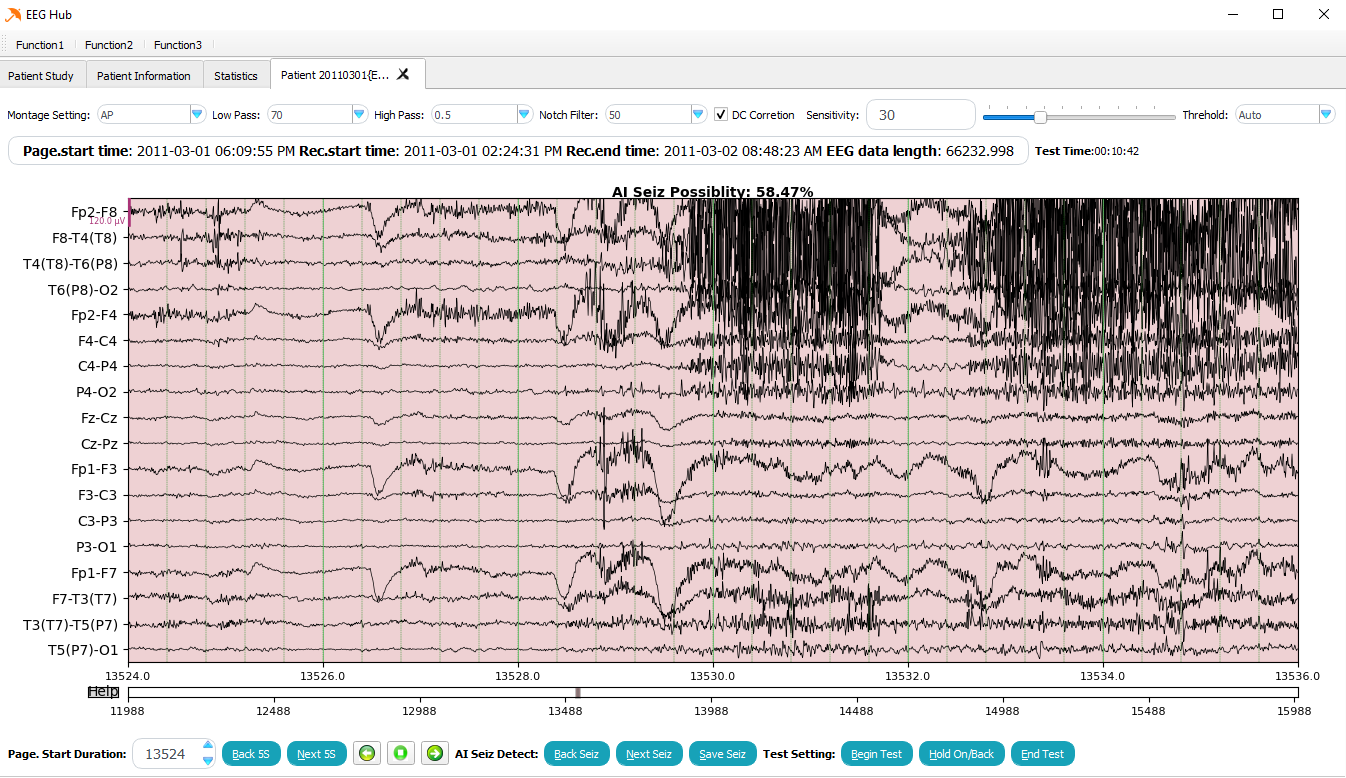}
    \caption{Our user interface (UI) used in clinical tests.}
    \label{fig:interface}
\vskip -.2cm
\end{figure*}

An example of the machine learning interface is shown in Fig.~\ref{fig:interface}. 

\subsection{Results}
The entire 1,142 sessions of RPAH dataset are labeled by a group of board-certified neurologists and EEG specialists over 9 years (2011-2019). Among these 1,142 sessions, there are 136 sessions with seizures that are only detected conclusively with a necessary video input. Hence our AI performance is measured on 1,006 sessions. This test is done in inference-only mode, and its performance is measured against a legacy ground truth. Our clinical test is performed with randomly selected 66 sessions with our AI-assisted system and our user interface (UI) tool with the participation of an expert human arbiter for review. The detailed results of our standalone AI system performance (on 1,006 EEG sessions) are shown in Table~\ref{tab:detail_result}. As explained in the paper, the 66 sessions in our pilot clinical study, out of a total 1,006, are reviewed and assessed by two neurologists and three EEG technicians to establish and validate a ground truth for our AI-assisted performance evaluation. Detail of this analysis is provided in Table~\ref{tab:clinical_test_detail}.

During the 66 sessions clinical test, as mentioned before, neurologists found five extra seizures with the help of the interface, and compared with the seizure that AI miss, three of them are confirmed can not directly find with only EEG information.
\begin{table*}[ht!]
    \centering
    \caption{Clinical test detail information}

\resizebox{0.75\textwidth}{!}
{

\begin{tabular}{cccccc}
\toprule
{\bf Session Number} & {\bf No. seizures (Human arbiter)} & {\bf No. seizures (AI+Human arbiter) } & {\bf Clinical test time} & {\bf Recording length} \\
\midrule
$1$ &$0$ & $0$&$0:07:03$ &$18:23:53$\\
$2$ &$3$&$3$&$0:05:28$ &$23:44:16$\\
$3$ &$8$&$9$&$0:11:07$ &$17:27:31$\\
$4$ &$3$&$4$&$0:01:02$ &$06:10:57$\\
$5$ &$0$ & $0$&$0:00:11$ &$16:27:02$\\
$6$ &$0$ & $0$&$0:05:54$ &$18:37:44$\\
$7$ &$0$ & $0$&$0:01:06$ &$18:15:15$\\
$8$ &$0$ & $0$&$0:02:24$ &$17:11:10$\\
$9$	&$0$ & $0$&$0:01:00$&$0:51:48$\\
$10$ & $0$&$0$&$0:03:12$&$14:04:13$\\
$11$&	$0$&	$0$&$	0:00:23$&$	10:18:34$\\
$12$&	$0$&	$0$&$	0:02:28$&$	19:51:37$\\
$13$&	$0$&	$0$&$	0:01:10$&$	06:40:56$\\
$14$&	$0$&	$0$&$	0:00:28$&$	06:21:43$\\
$15$&	$0$&	$0$&$	0:05:13$&$	06:50:31$\\
$16$&	$0$&	$0$&$	0:03:34$&$	25:23:21$\\
$17$&	$1$&	$1$&$	0:03:21$&$	02:46:42$\\
$18$&	$1$&	$1$&$	0:07:31$&$	23:41:59$\\
$19$&	$0$&	$0$&$	0:09:25$&$	17:00:39$\\
$20$&	$2$&	$2$&$	0:08:37$&$	24:01:36$\\
$21$&	$1$&	$1$&$	0:02:54$&$	24:42:02$\\
$22$&	$0$&	$0$&$	0:09:54$&$	20:41:51$\\
$23$&	$1$&	$1$&$	0:00:33$&$	01:25:52$\\
$24$&	$1$&	$1$&$	0:11:09$&$	30:16:09$\\
$25$&	$0$&	$0$&$	0:00:25$&$	02:24:13$\\
$26$&	$0$&	$0$&$	0:03:35$&$	20:31:00$\\
$27$&	$0$&	$0$&$	0:04:30$&$	16:19:59$\\
$28$&	$0$&	$0$&$	0:00:29$&$	01:24:35$\\
$29$&	$0$&	$0$&$	0:00:03$&$	25:49:53$\\
$30$&	$0$&	$0$&$	0:00:31$&$	01:19:17$\\
$31$&	$0$&	$0$&$	0:00:53$&$	03:10:56$\\
$32$&	$0$&	$0$&$	0:07:41$&$	24:12:53$\\
$33$&	$1$&	$1$&$	0:10:58$&$	04:55:00$\\
$34$&	$0$&	$0$&$	0:12:20$&$	23:47:51$\\
$35$&	$2$&	$3$&$	0:02:37$&$	21:26:48$\\
$36$&	$0$&	$0$&$	0:09:08$&$	21:39:24$\\
$37$&	$0$&	$0$&$	0:02:24$&$	20:18:14$\\
$38$&	$0$&	$0$&$	0:04:44$&$	09:00:22$\\
$39$&	$0$&	$0$&$	0:04:41$&$	23:52:12$\\
$40$&	$0$&	$0$&$	0:04:47$&$	14:53:24$\\
$41$&	$0$&	$0$&$	0:00:10$&$	00:25:32$\\
$42$&	$0$&	$0$&$	0:05:23$&$	06:26:45$\\
$43$&	$3$&	$3$&$	0:05:24$&$	17:00:31$\\
$44$&	$0$&	$0$&$	0:00:46$&$	02:49:18$\\
$45$&	$0$&	$0$&$	0:00:09$&$	01:19:10$\\
$46$&	$0$&	$0$&$	0:02:18$&$	01:33:55$\\
$47$&	$1$&	$1$&$	0:13:09$&$	23:51:00$\\
$48$&	$1$&	$1$&$	0:11:15$&$	18:44:52$\\
$49$&	$0$&	$0$&$	0:06:33$&$	23:01:36$\\
$50$&	$0$&	$0$&$	0:02:02$&$	06:23:43$\\
$51$&	$0$&	$0$&$	0:00:46$&$	00:29:46$\\
$52$&	$1$&	$1$&$	0:02:13$&$	00:56:00$\\
$53$&	$2$&	$2$&$	0:08:08$&$	08:04:06$\\
$54$&	$17$&	$15$&$	0:11:17$&$	24:16:04$\\
$55$&	$3$  &$2$&	$0:11:37$&$	15:10:56$\\
$56$&	$0$&	$0$&$	0:02:15$&$	06:03:09$\\
$57$&	$0$&	$0$&$	0:03:47$&$	16:13:56$\\
$58$&	$0$&	$0$&$	0:01:05$&$	05:42:41$\\
$59$&	$2$&	$2$&$	0:04:29$&$	14:37:38$\\
$60$&	$4$&	$4$&$	0:07:54$&$	15:27:53$\\
$61$&	$1$&	$1$&$	0:00:46$&$	08:27:35$\\
$62$&	$0$&	$0$&$	0:02:46$&$	16:13:24$\\
$63$&	$0$&	$0$&$	0:00:54$&$	16:13:55$\\
$64$&	$0$&	$0$&$	0:00:53$&$	07:40:35$\\
$65$&	$0$&	$0$&$	0:00:18$&$	06:58:55$\\
$66$&	$0$&	$0$&$	0:01:04$&$	18:38:22$\\
\midrule
Overall &$59$&$59$&$4:42:14$&$889:14:39$ \\
\bottomrule
\end{tabular}
}

\label{tab:clinical_test_detail}
\vskip -.2cm
\end{table*}

\begin{table*}[ht!]
    \centering
    \caption{Details of RPAH annual results}
    \label{tab:detail_result}
    \resizebox{0.9\textwidth}{!}
    {
        \begin{tabular}{lccccccccc}  
         
            \toprule
            Year&
            AUC & 
            SDR&
            FA/$24$~hrs& 
            SDR*&
            FA/$24$~hrs*& 
            Total & 
            Number of &
            Number of&
            Number of 
            \\
             &
             & 
            &
            & 
            &
            & 
            duration~(hours) & 
            sessions &
            patients &
            patients with seizure
            \\
            \midrule
            $2011$&$0.8993$&  $82.45\%$  & $58.00$&  $85.96\%$  & $57.96$ & $1114.69$&
            $75$&$11$&$9$\\
            
            $2012$&$0.9107$&  $83.52\%$  & $56.77$ &  $83.52\%$  & $56.76$& $1752.62$&
            $117$&$23$&$14$\\
            
            $2013$&$0.896$&  $83.33\%$  & $47.11$&  $86.67\%$  & $47.09$ & $2090.99$&
            $118$&$22$&$9$\\
            
            $2014$&$0.7382$&  $73.02\%$  & $74.86$&  $74.60\%$  & $74.84$ & $1792.13$&
            $111$&$24$&$12$\\
            
            $2015$&$0.8215$&  $78.69\%$  & $64.78$&  $78.69\%$  & $64.74$ & $2075.35$&
            $139$&$28$&$16$\\
            
            $2016$&$0.8547$&  $80.25\%$  & $44.75$&  $80.25\%$  & $44.72$ & $1506.03$&
            $101$&$19$&$14$\\
            
            $2017$&$0.6827$&  $67.65\%$  & $53.41$&  $70.58\%$  & $53.36$ & $2171.21$&
            $174$&$24$&$16$\\
            
            $2018$&$0.7286$&  $58.49\%$  & $50.17$&  $66.04\%$  & $50.07$ & $1181.05$&
            $100$&$20$&$15$\\
            
            $2019$&$0.7877$&  $75.00\%$  & $56.54$&  $75.00\%$  & $56.48$ & $907.60$&
            $71$&$21$&$6$\\
            \midrule
            Overall&$0.8172$&  $76.68\%$  & $56.55$&  $78.54\%$  & $56.52$ & $14591.6$&
            $1006$&$192$&$111$\\
            \bottomrule   
        \end{tabular}    
        }

        \begin{tablenotes}
            \item[] {\scriptsize Note: The metric AUC, SDR, FA/24 hours are explained in detail in the Main Body metrics section. SDR* and FA/24 hours* method combines all false alarms within 30-seconds as one.}
        \end{tablenotes}

\end{table*}

\subsubsection{Extra seizures detected by AI}
\begin{figure*}[ht!]

\centering
    \includegraphics[width=1.0\textwidth]{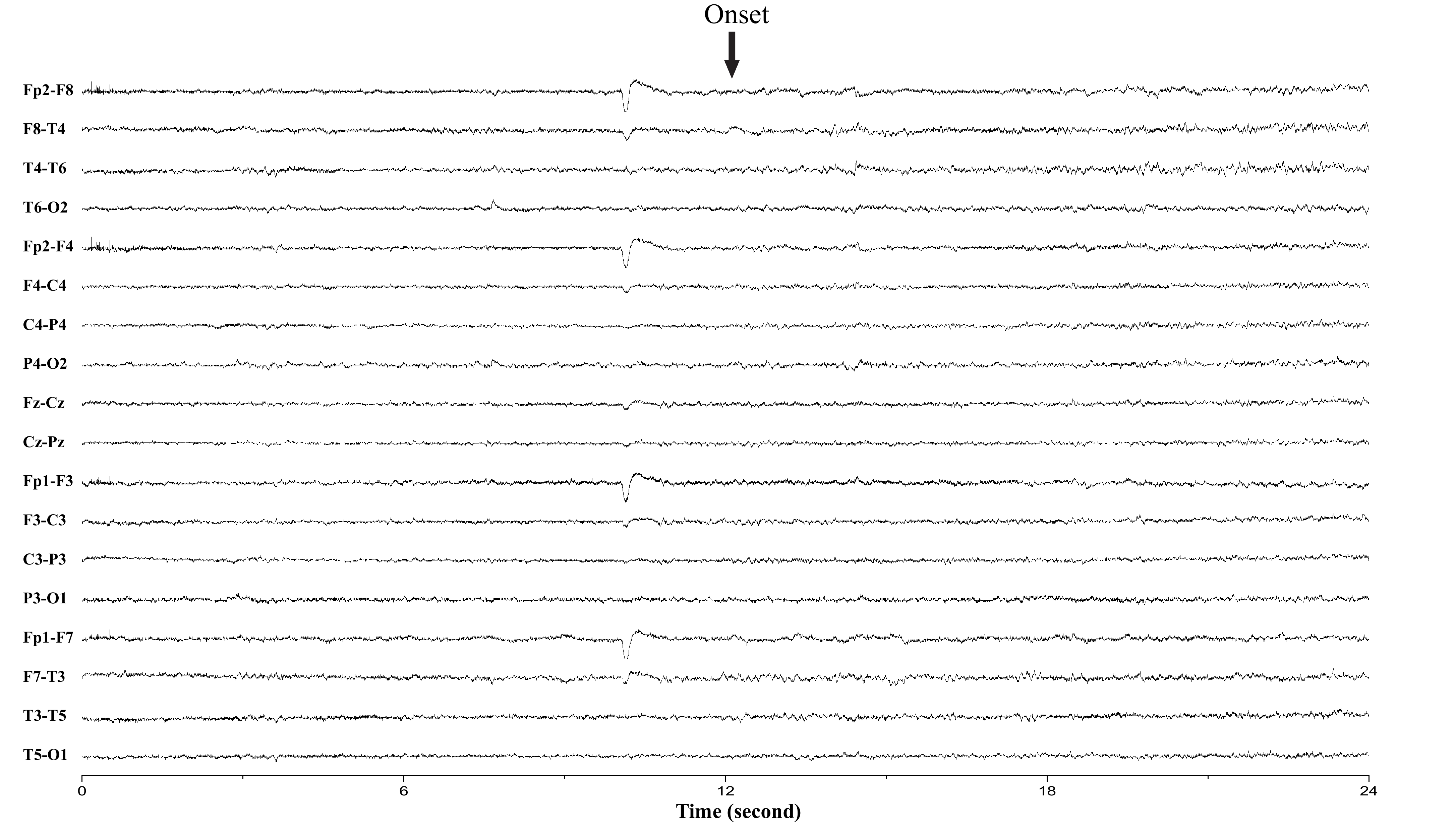}
   \caption{Extra seizure detected by AI (verified by the neurologist).}
    \label{fig:AI_extra1}

\end{figure*}

\begin{figure*}[ht!]

\centering
    \includegraphics[width=1.0\textwidth]{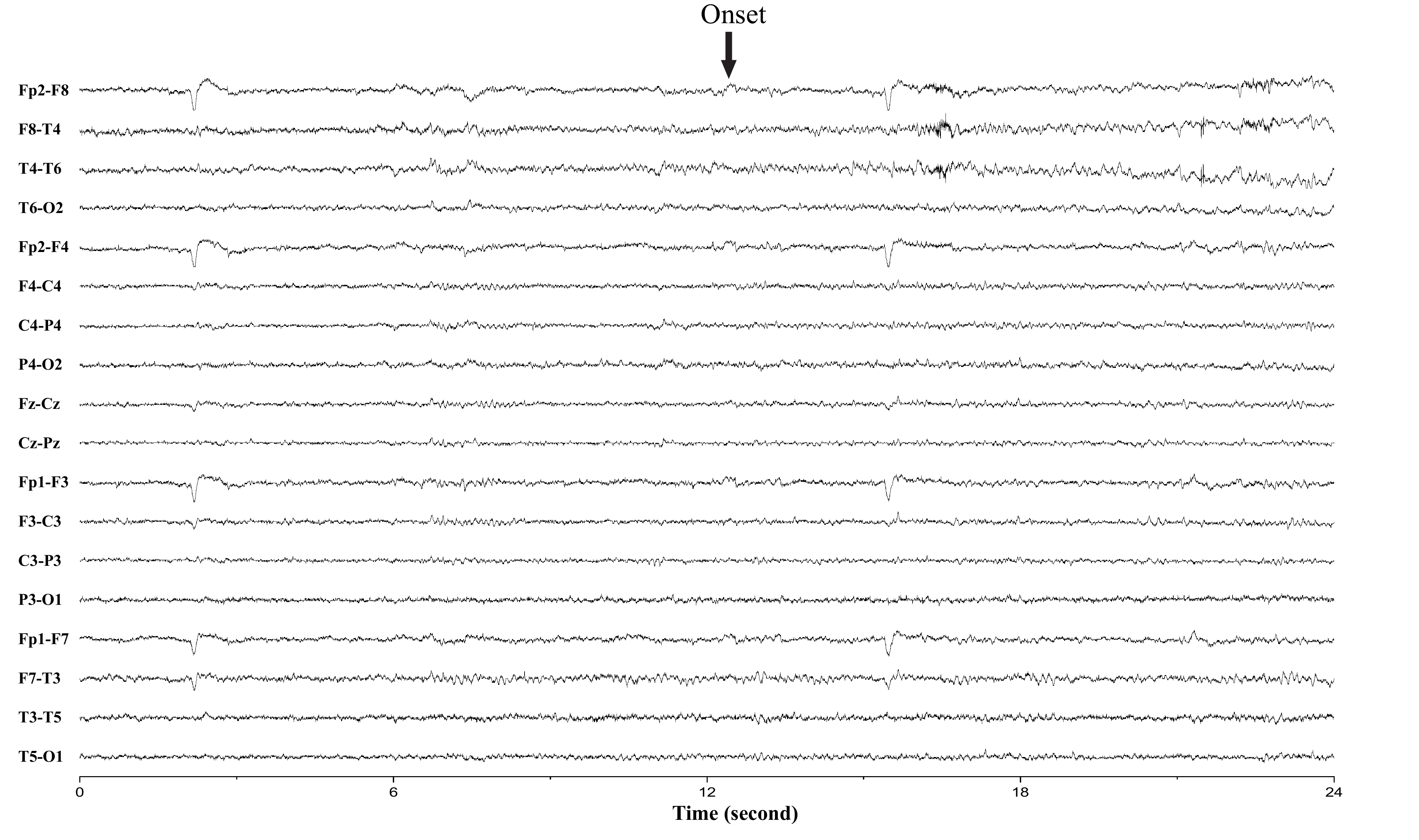}
   \caption{Extra seizure detected by AI (verified by the neurologist).}
    \label{fig:AI_extra3}

\end{figure*}

\begin{figure*}[ht!]

\centering
    \includegraphics[width=1.0\textwidth]{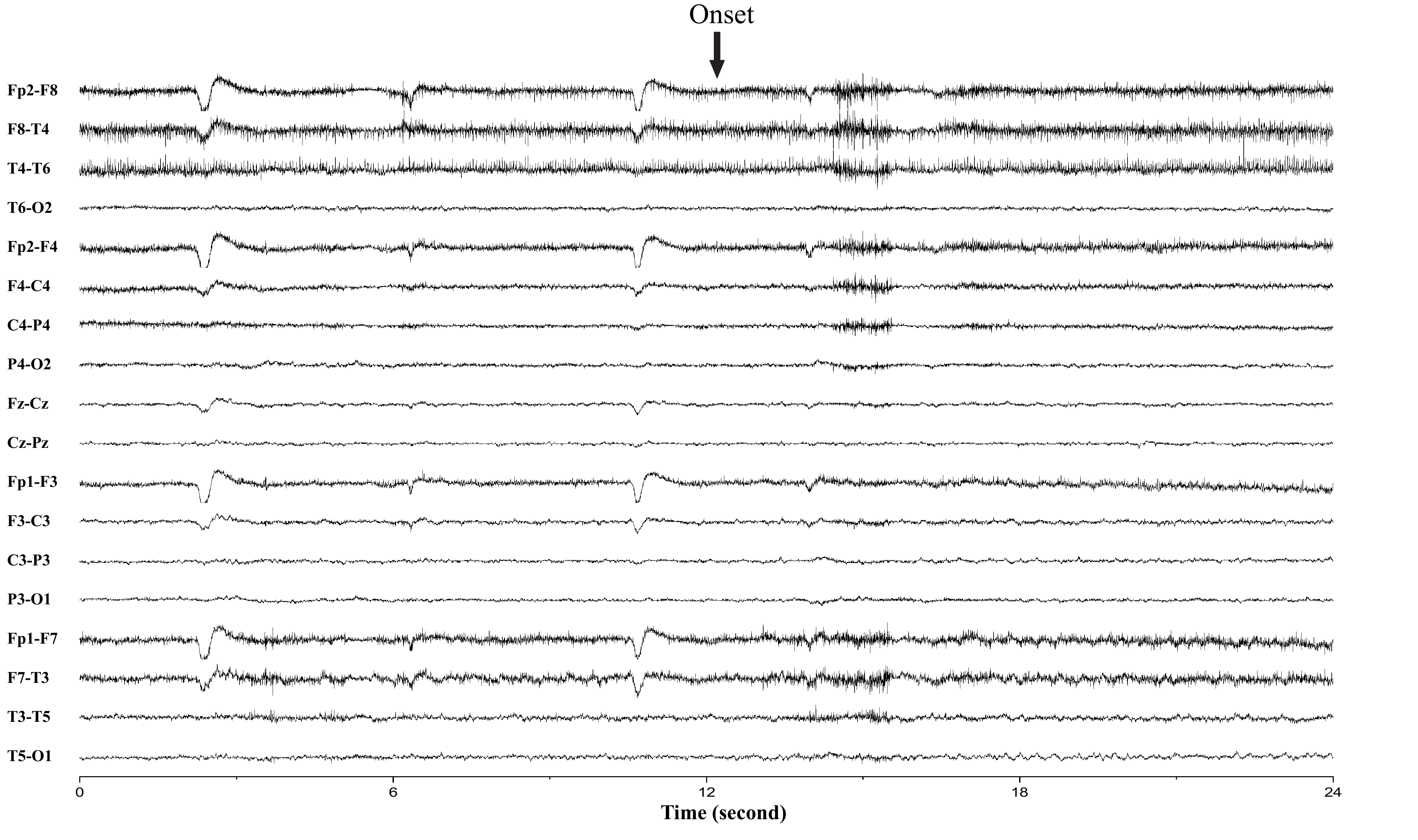}
   \caption{Extra seizure detected by AI (verified by the neurologist).}
    \label{fig:AI_extra9}

\end{figure*}

\begin{figure*}[ht!]

\centering
    \includegraphics[width=1.0\textwidth]{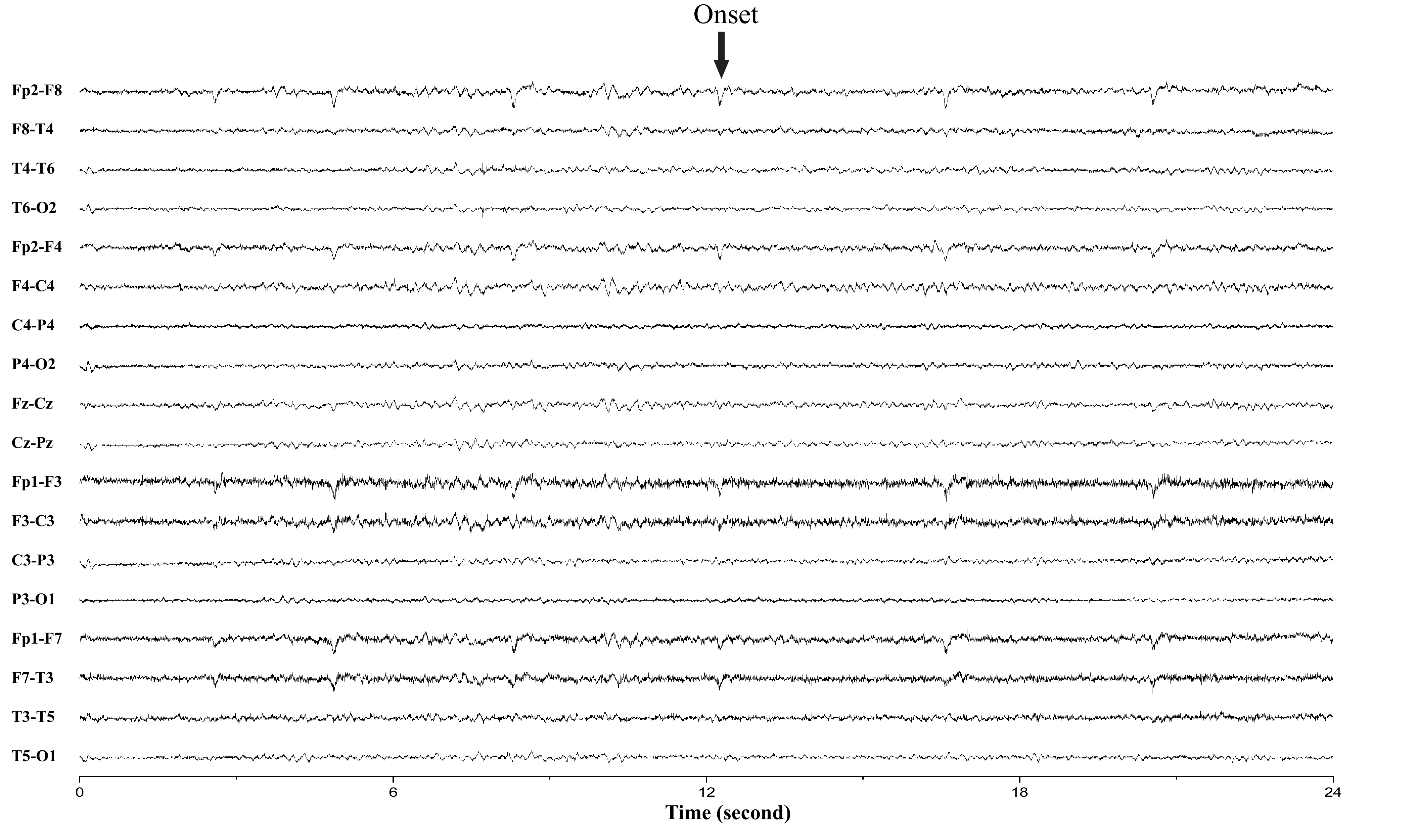}
   \caption{Extra seizure detected by AI (verified by the neurologist).}
    \label{fig:AI_extra14}

\end{figure*}

The examples (one in the main body) are shown in Fig~\ref{fig:AI_extra1}, \ref{fig:AI_extra3}, \ref{fig:AI_extra9}, \ref{fig:AI_extra14}

\subsubsection{AI miss seizures}
\begin{figure*}[ht!]
\centering
    \includegraphics[width=1.0\textwidth]{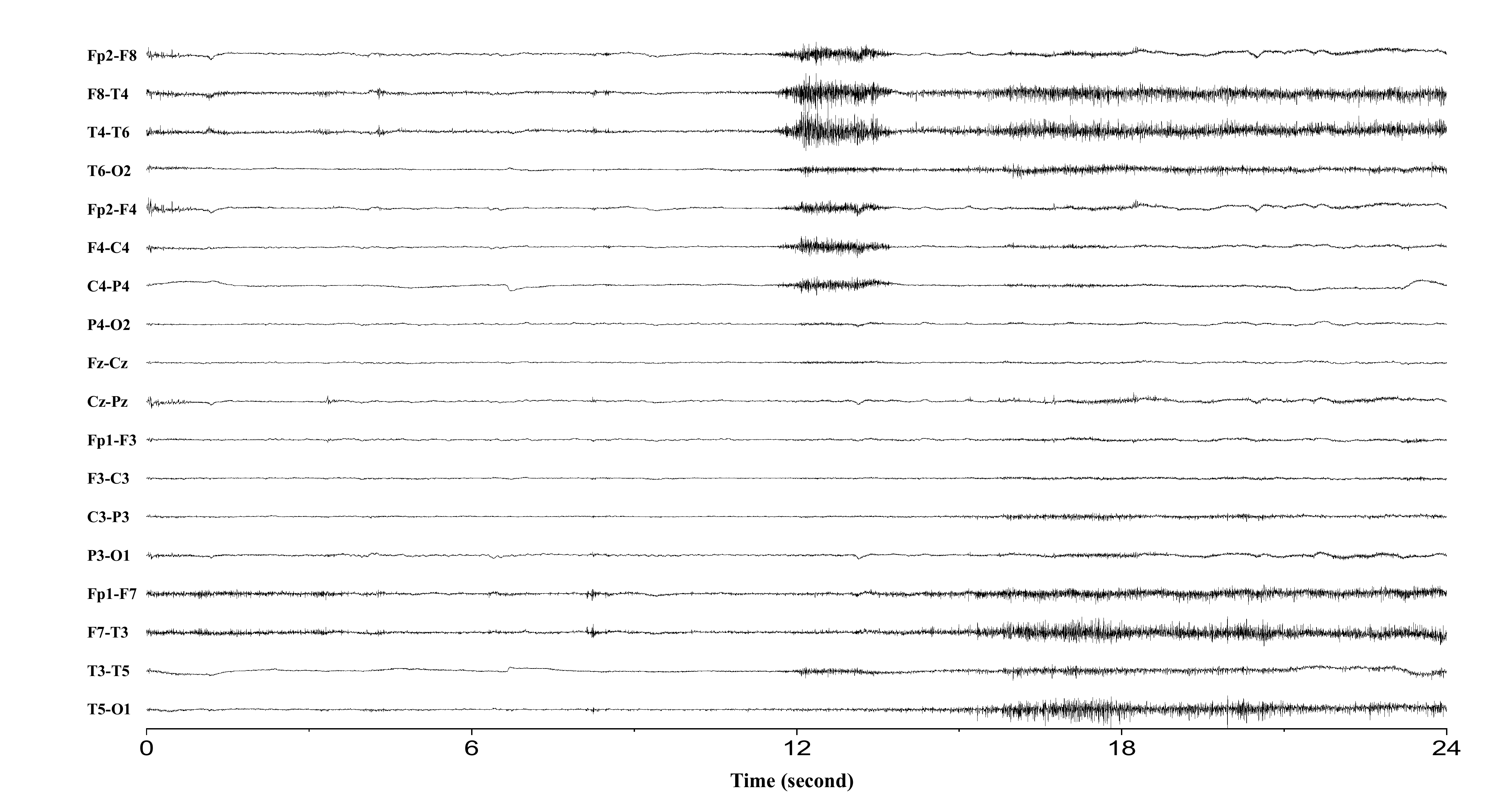}
    \caption{Seizure undetected by AI. There is no clear frequency evolution, and there are lots of muscle and eye artifacts that can only be confirmed by the video (verified by the neurologist).}
    \label{fig:AI_miss2}

\end{figure*}

\begin{figure*}[ht!]

\centering
    \includegraphics[width=1.0\textwidth]{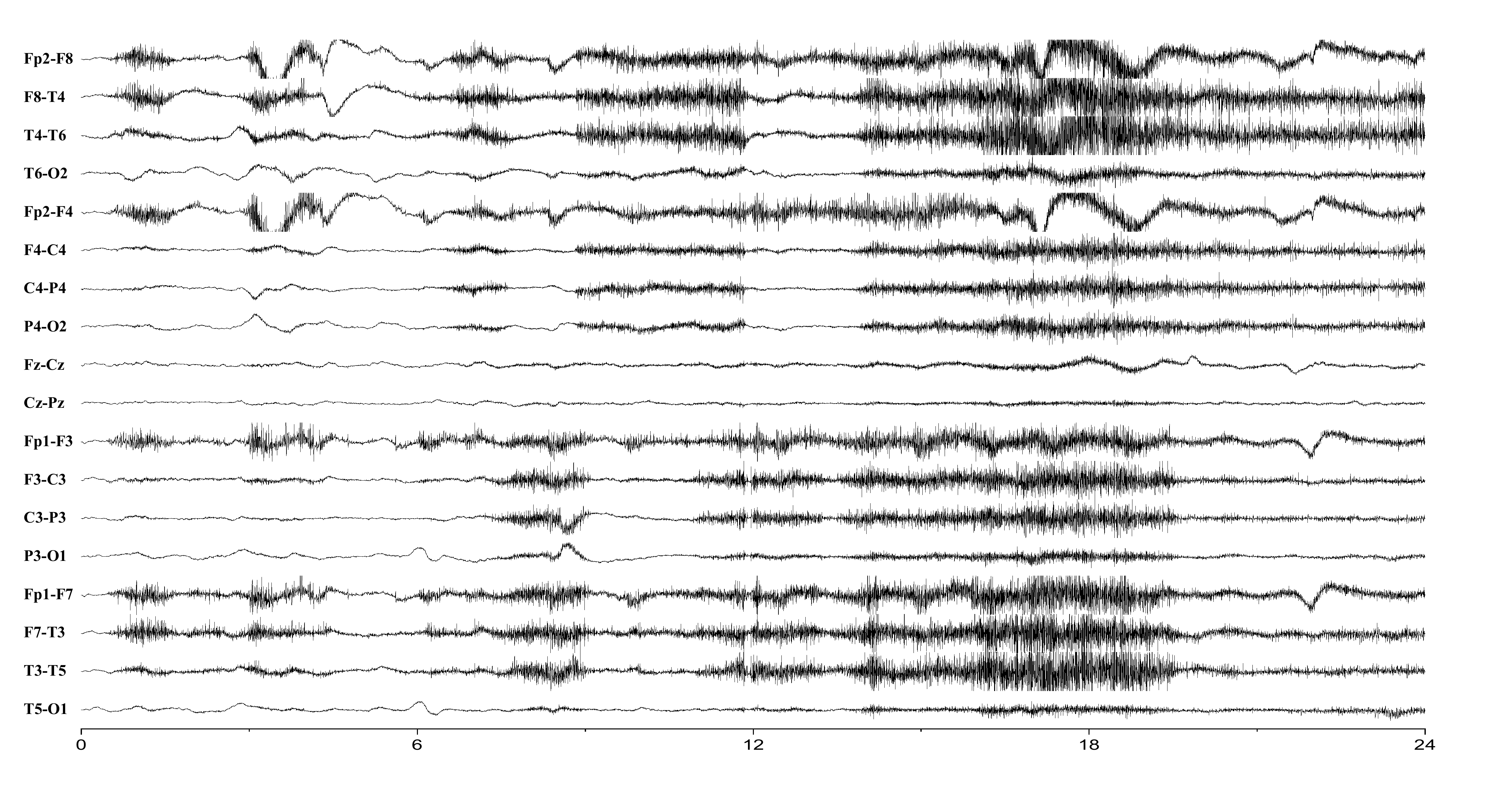}
   \caption{Seizure undetected by AI. There is no clear frequency evolution, and there are lots of muscle artifacts that can only be confirmed by the video (verified by the neurologist).}
    \label{fig:AI_miss3}

\end{figure*}

\begin{figure*}[ht!]
\centering
    \includegraphics[width=1.0\textwidth]{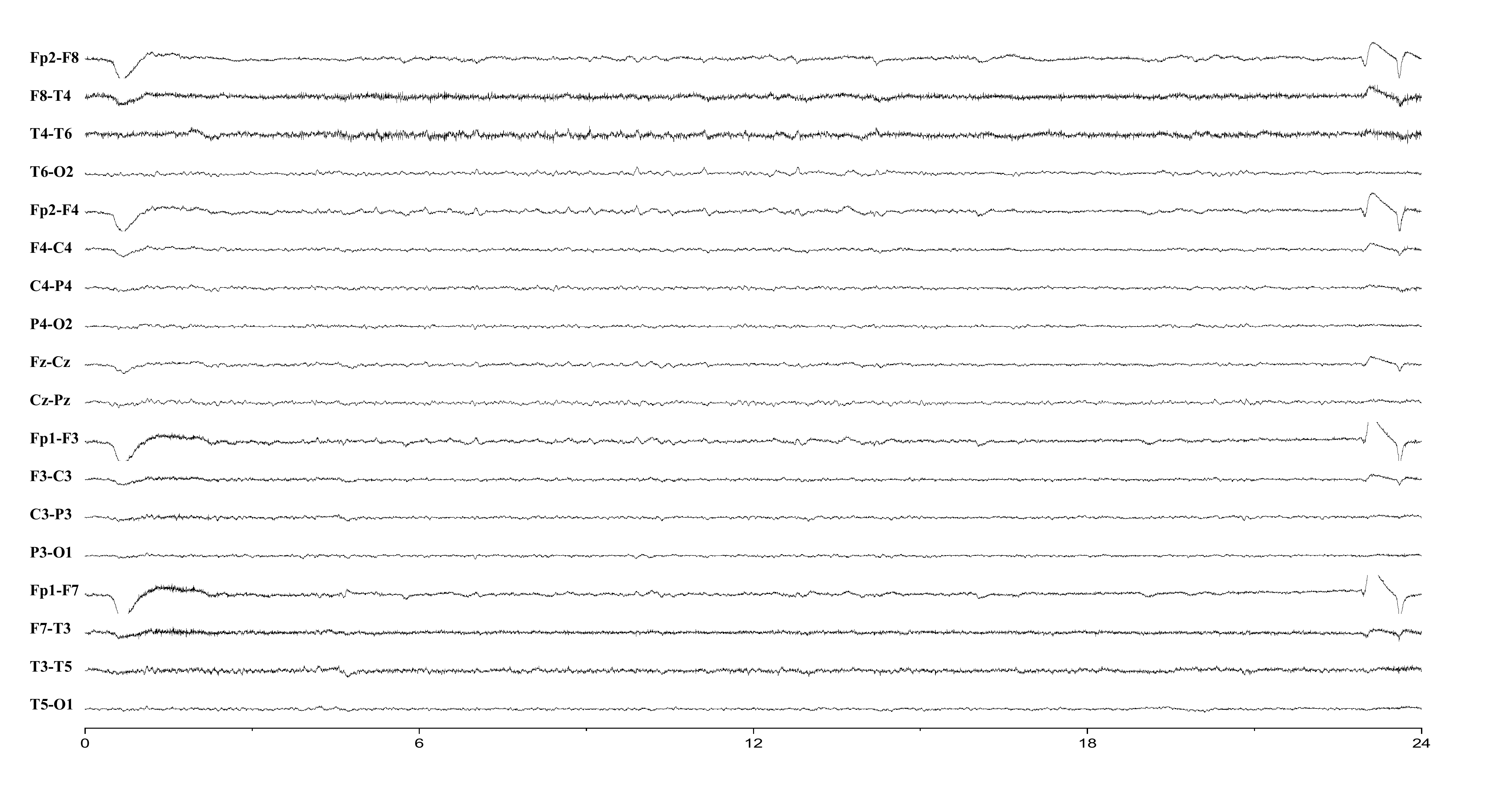}
   \caption{Seizure undetected by AI. There is no clear frequency evolution, and there are lots of muscle artifacts that can only be confirmed by the video (verified by the neurologist).}
    \label{fig:AI_miss4}

\end{figure*}

Three examples are shown in Fig~\ref{fig:AI_miss2}, \ref{fig:AI_miss3}, and \ref{fig:AI_miss4}, which are confirmed only on video and not but EEG information.

\begin{thebibliography}{10}

\bibitem{haibe2020transparency}
Benjamin Haibe-Kains, George~Alexandru Adam, Ahmed Hosny, Farnoosh Khodakarami,
  Levi Waldron, Bo~Wang, Chris McIntosh, Anna Goldenberg, Anshul Kundaje,
  Casey~S Greene, et~al.
\newblock Transparency and reproducibility in artificial intelligence.
\newblock {\em Nature}, 586(7829):E14--E16, 2020.

\bibitem{kelly2019key}
Christopher~J Kelly, Alan Karthikesalingam, Mustafa Suleyman, Greg Corrado, and
  Dominic King.
\newblock Key challenges for delivering clinical impact with artificial
  intelligence.
\newblock {\em {BMC Medicine}}, 17(1):195, 2019.

\bibitem{mckinney2020international}
Scott~Mayer McKinney, Marcin Sieniek, Varun Godbole, Jonathan Godwin, Natasha
  Antropova, Hutan Ashrafian, Trevor Back, Mary Chesus, Greg~S Corrado, Ara
  Darzi, et~al.
\newblock International evaluation of an {AI} system for breast cancer
  screening.
\newblock {\em Nature}, 577(7788):89--94, 2020.

\bibitem{baumgartner2018seizure}
Christoph Baumgartner and Johannes~P Koren.
\newblock Seizure detection using scalp-eeg.
\newblock {\em Epilepsia}, 59:14--22, 2018.

\bibitem{koren2021systematic}
Johannes Koren, Sebastian Hafner, Moritz Feigl, and Christoph Baumgartner.
\newblock Systematic analysis and comparison of commercial seizure-detection
  software.
\newblock {\em Epilepsia}, 62(2):426--438, 2021.

\bibitem{shoeibi2020epileptic}
Afshin Shoeibi, Navid Ghassemi, Marjane Khodatars, Mahboobeh Jafari, Sadiq
  Hussain, Roohallah Alizadehsani, Parisa Moridian, Abbas Khosravi, Hossein
  Hosseini-Nejad, Modjtaba Rouhani, et~al.
\newblock Epileptic seizure detection using deep learning techniques: a review.
\newblock {\em arXiv preprint arXiv:2007.01276}, 2020.

\bibitem{abend2015much}
Nicholas~S Abend, Alexis~A Topjian, and Sankey Williams.
\newblock How much does it cost to identify a critically ill child experiencing
  electrographic seizures?
\newblock {\em Journal of Clinical Neurophysiology}, 32(3):257, 2015.

\bibitem{golmohammadi2020deep}
Meysam Golmohammadi, Vinit Shah, Iyad Obeid, and Joseph Picone.
\newblock Deep learning approaches for automated seizure detection from scalp
  electroencephalograms.
\newblock In {\em Signal Processing in Medicine and Biology}, pages 235--276.
  Springer, 2020.

\bibitem{saab2020weak}
Khaled Saab, Jared Dunnmon, Christopher R{\'e}, Daniel Rubin, and Christopher
  Lee-Messer.
\newblock Weak supervision as an efficient approach for automated seizure
  detection in electroencephalography.
\newblock {\em npj Digital Medicine}, 3(1):1--12, 2020.

\bibitem{zech2018variable}
John~R Zech, Marcus~A Badgeley, Manway Liu, Anthony~B Costa, Joseph~J Titano,
  and Eric~Karl Oermann.
\newblock Variable generalization performance of a deep learning model to
  detect pneumonia in chest radiographs: a cross-sectional study.
\newblock {\em PLoS Medicine}, 15(11):e1002683, 2018.

\bibitem{furbass2015prospective}
F~F{\"u}rbass, P~Ossenblok, M~Hartmann, H~Perko, AM~Skupch, G~Lindinger,
  L~Elezi, E~Pataraia, AJ~Colon, C~Baumgartner, et~al.
\newblock Prospective multi-center study of an automatic online seizure
  detection system for epilepsy monitoring units.
\newblock {\em Clinical Neurophysiology}, 126(6):1124--1131, 2015.

\bibitem{scheuer2020seizure}
Mark~L Scheuer, Scott~B Wilson, Arun Antony, Gena Ghearing, Alexandra Urban,
  and Anto~I Bagi{\'c}.
\newblock Seizure detection: interreader agreement and detection algorithm
  assessments using a large dataset.
\newblock {\em Journal of Clinical Neurophysiology}, 00(00), 2020.

\bibitem{furbass2018eeg}
Franz F{\"u}rbass.
\newblock {\em {EEG} monitoring based on automatic detection of seizures and
  repetitive discharges}.
\newblock PhD thesis, Technische Universität Wien, 2018.

\bibitem{xingjian2015convolutional}
SHI Xingjian, Zhourong Chen, Hao Wang, Dit-Yan Yeung, Wai-Kin Wong, and
  Wang-chun Woo.
\newblock {Convolutional {LSTM} network: a machine learning approach for
  precipitation nowcasting}.
\newblock {\em Advances in Neural Information Processing Systems}, pages
  802--810, 2015.

\bibitem{hartmann2011episcan}
Manfred~M Hartmann, Franz F{\"u}rba{\ss}, Hannes Perko, Ana Skupch, Katharina
  Lackmayer, Christoph Baumgartner, and Tilmann Kluge.
\newblock {EpiScan}: online seizure detection for epilepsy monitoring units.
\newblock {\em Proc. IEEE Engineering in Medicine and Biology Society}, pages
  6096--6099, 2011.

\bibitem{shah2018temple}
Vinit Shah, Eva Von~Weltin, Silvia Lopez, James~Riley McHugh, Lillian Veloso,
  Meysam Golmohammadi, Iyad Obeid, and Joseph Picone.
\newblock The temple university hospital seizure detection corpus.
\newblock {\em Frontiers in Neuroinformatics}, 12:83, 2018.

\bibitem{fisher1992high}
Robert~S Fisher, WR~Webber, Ronald~P Lesser, Santiago Arroyo, and Sumio
  Uematsu.
\newblock High-frequency {EEG} activity at the start of seizures.
\newblock {\em Journal of Clinical Neurophysiology}, 9(3):441--448, 1992.

\bibitem{comon1994independent}
Pierre Comon.
\newblock Independent component analysis, a new concept?
\newblock {\em Signal Processing}, 36(3):287--314, 1994.

\bibitem{belouchrani1997blind}
Adel Belouchrani, Karim Abed-Meraim, J-F Cardoso, and Eric Moulines.
\newblock A blind source separation technique using second-order statistics.
\newblock {\em IEEE Transactions on Signal Processing}, 45(2):434--444, 1997.

\bibitem{dammers2008integration}
Jurgen Dammers, Michael Schiek, Frank Boers, Carmen Silex, Mikhail Zvyagintsev,
  Uwe Pietrzyk, and Klaus Mathiak.
\newblock Integration of amplitude and phase statistics for complete artifact
  removal in independent components of neuromagnetic recordings.
\newblock {\em IEEE Transactions on Biomedical Engineering}, 55(10):2353--2362,
  2008.

\bibitem{gramfort2013meg}
Alexandre Gramfort, Martin Luessi, Eric Larson, Denis~A Engemann, Daniel
  Strohmeier, Christian Brodbeck, Roman Goj, Mainak Jas, Teon Brooks, Lauri
  Parkkonen, et~al.
\newblock {MEG} and {EEG} data analysis with {MNE}-{Python}.
\newblock {\em Frontiers in Neuroscience}, 7:267, 2013.

\bibitem{shah2017optimizing}
Vinit Shah, Meysam Golmohammadi, Saeedeh Ziyabari, Eva Von~Weltin, Iyad Obeid,
  and Joseph Picone.
\newblock Optimizing channel selection for seizure detection.
\newblock {\em Proc. IEEE Signal Processing in Medicine and Biology Symposium
  (SPMB)}, pages 1--5, 2017.

\bibitem{ziyabari2017objective}
Saeedeh Ziyabari, Vinit Shah, Meysam Golmohammadi, Iyad Obeid, and Joseph
  Picone.
\newblock {Objective evaluation metrics for automatic classification of {EEG}
  events}.
\newblock {\em arXiv preprint arXiv:1712.10107}, 2017.

\bibitem{roy2021evaluation}
Subhrajit Roy, Isabell Kiral-Kornek, Mahtab Mirmomeni, Todd Mummert, Alan Braz,
  Jason Tsai, Jianbin Tang, Umar Asif, Thomas Schaffter, Mehmet~Eren Ahsen10,
  et~al.
\newblock Evaluation of combined artificial intelligence and neurologist
  assessment to annotate scalp electroencephalography data.
\newblock {\em EBioMedicine}, 103275, 2021.

\bibitem{pavel2020machine}
Andreea~M Pavel, Janet~M Rennie, Linda~S de~Vries, Mats Blennow, Adrienne
  Foran, Divyen~K Shah, Ronit~M Pressler, Olga Kapellou, Eugene~M Dempsey,
  Sean~R Mathieson, et~al.
\newblock A machine-learning algorithm for neonatal seizure recognition: a
  multicentre, randomised, controlled trial.
\newblock {\em The Lancet Child \& Adolescent Health}, 4(10):740--749, 2020.

\bibitem{goetz2007textbook}
Nancy Foldvary‐Schaefer and Elaine Wyllie.
\newblock {Chapter 52 - Epilepsy}.
\newblock In Christopher~G. Goetz, editor, {\em Textbook of Clinical
  Neurology}, pages 1213--1244. W.B. Saunders, Philadelphia, 3rd edition, 2007.

\end{thebibliography}
\end{document}